\documentclass[twocolumn]{aastex62}
\usepackage{graphicx}

\def\npar {\partial}

\submitjournal{ApJ}

\shorttitle{Plasma Heating in the Current Sheet in a 3D CME Simulation} 
\shortauthors{Reeves et al.} 

\begin{document}

\title{Exploring Plasma Heating in the Current Sheet Region in a Three-Dimensional Coronal Mass Ejection Simulation}
\author[0000-0002-6903-6832]{Katharine K. Reeves}
\email{{\tt kreeves@cfa.harvard.edu}}
\affil{Harvard-Smithsonian Center for Astrophysics, 60 Garden St. MS 58, Cambridge, MA 02138}
\author[0000-0003-3843-3242]{Tibor T\"{o}r\"{o}k}
\author[0000-0002-3164-930X]{Zoran Miki\'{c}}
\author{Jon Linker} 
\affil{Predictive Science, Inc. (PSI), 9990 Mesa Rim Road, Suite 170, San Diego, CA 92121-2910}
\author[0000-0001-6628-8033]{Nicholas A. Murphy}
\affil{Harvard-Smithsonian Center for Astrophysics, 60 Garden St. MS 58, Cambridge, MA 02138}

\received{July 24, 2019}
\revised{October 8, 2019}

\begin{abstract}
We simulate a coronal mass ejection (CME) using a three-dimensional magnetohydrodynamic (MHD) code that includes coronal heating, thermal conduction, and radiative cooling in the energy equation. The magnetic flux distribution at 1 R$_s$ is produced by a localized subsurface dipole superimposed on a global dipole field, mimicking the presence of an active region within the global corona.  
Transverse electric fields are applied near the polarity inversion line to introduce a transverse magnetic field, followed by the imposition of a converging flow to form and destabilize a flux rope, producing an eruption.
We examine the quantities responsible for plasma heating and cooling during the eruption, including thermal conduction, radiation, adiabatic effects, coronal heating, and ohmic heating.  We find that ohmic heating is an important contributor to hot temperatures in the current sheet region early in the eruption, but in the late phase adiabatic compression plays an important role in heating the plasma there. Thermal conduction also plays an important role in the transport of thermal energy away from the current sheet region throughout the reconnection process, producing a ``thermal halo'' and widening the region of high temperatures.  We simulate emission from solar telescopes for this eruption and find that there is evidence for emission from heated plasma above the flare loops late in the eruption, when the adiabatic heating is the dominant heating term.  These results provide an explanation for hot supra-arcade plasma sheets that are often observed in X-rays and extreme ultraviolet wavelengths during the decay phase of large flares.
 \end{abstract}

\section{Introduction}

Solar eruptions such as flares and CMEs are dynamic events responsible for the release of energy that can be on the order of $10^{32}$ ergs \citep[e.g.,][]{Forbes2000}.  This energy is stored in the form of magnetic energy prior to the eruption due to stresses in the Sun's magnetic field.  During the eruption, this stored energy is converted to thermal energy of the heated plasma, non-thermal kinetic energy that accelerates particles, and the bulk kinetic energy of the resulting CME.   How this energy is converted and transported, resulting in the observable emission from the Sun during solar 
eruptions, is an active area of current research.  

An observable consequence of the energy release in solar eruptions is the heating of the plasma in the current sheet region and in the reconnected flare loops that form below an erupting CME.  Hot plasma sheets have been observed above flare arcades (often referred to as ``supra-arcade'' plasma sheets) in X-rays and extreme ultraviolet (EUV) wavelengths \citep{McKenzie1999, Savage2012} with temperatures on the order of 10 MK \citep{Innes_b2003, Reeves2011, Hanneman2014, Reeves2017, Warren2018}.    These plasma sheets are sometimes seen connecting the flare loops and the erupting CME \citep[e.g.][]{Reeves2011, Yan2018}, and they are believed to be related to the current sheet formed during the eruption of a magnetic flux rope.  These structures persist for long times in the late phase of an eruption \citep[e.g.][]{Hanneman2014,Savage2012}, lasting much longer than the conductive cooling time \citep{Reeves2017}. This finding indicates that there must either be some heating mechanism in the vicinity of the current sheet that keeps the plasma from cooling, or a suppression of the thermal conduction in this region (or both).  Additionally, spectroscopic observations demonstrate that the plasma in these regions have coronal abundances, as opposed to the photospheric abundances that exist in the flare loops \citep{Landi2012,Warren2018}, indicating that supra-arcade plasma is coronal in origin.   

A handful of numerical simulations have been performed with the goal of understanding how heating in the flare and current sheet region occurs. Work by \citet{YokoyamaShibata1997,YokoyamaShibata2001} and \citet{Chen1999} using two-dimensional (2D) flare simulations find that the addition of thermal conduction into the simulation broadened the area of high temperature around the current sheet.  This effect has also been seen in an analytic treatment of the current sheet region \citep{Seaton2009} and a 2.5D global CME simulation \citep{Reeves2010}.  The simulations by \citet{YokoyamaShibata1998,YokoyamaShibata2001} also show that thermal conduction can cause energy to be deposited in the chromosphere, creating a pressure imbalance and driving plasma into the flare loops, a process commonly referred to as ``chromospheric evaporation.''

Several authors have used 2D or 2.5D simulations to investigate how the energy is partitioned in the reconnection region during an eruption.  \citet{Magara1996} find that magnetic energy is efficiently converted into bulk kinetic energy of the erupting plasmoid, especially in a case approximating efficient thermal conduction.  A similar result was found by \citet{Hirose2001}, who determined that most of the magnetic energy in their simulation is converted into kinetic energy of the CME via the Lorentz force, and the heating of the flare loops is achieved via adiabatic plasma compression.  Ohmic dissipation does not contribute a large amount to the increase in internal energy in the simulation because it occurs in a small region.  On the other hand, \citet{Ugai2007} found that Ohmic heating can be responsible for heating the chromosphere enough to drive chromospheric evaporation.   In a simulation that includes thermal conduction and radiation in the energy equation, \citet{Reeves2010} find that the bulk of the energy flow in the direction of the flare loops is in the form of thermal conduction flux, and the energy flow in the direction of the CME is the Poynting flux associated with the azimuthal magnetic field early in the eruption, and kinetic energy flux at later times. 

\citet{Birn2009} use a 3D MHD simulation with a simplified energy equation to follow the energy conversion in a flare. Considering stretched arcade-like magnetic configurations, they trigger flare reconnection by imposing localized finite resistivity within the current sheet. They find that below the reconnection site, the incoming Poynting flux is largely converted into enthalpy flux, and in the CME direction it is largely converted to kinetic energy flux.  The energy of the outgoing enthalpy flux in the downward case is produced by a two-step process where the plasma is accelerated by the Lorentz force and then decelerated by pressure gradients, resulting in adiabatic heating in layers that extend along the current sheet.  They also find that there is only a small increase of ohmic dissipation during the fast energy release in the eruption, indicating that the ohmic heating plays a relatively minor role in the heat transport and dissipation in this simulation.

To our knowledge, no thorough accounting of the energy release and transport has been done in a three-dimensional (3D) simulation that includes a realistic energy equation and a self-consistent modeling of current-sheet formation and flare reconnection.  In this paper, we describe a simulation of a CME and the associated flare with a 3D code that includes conduction, radiation, and coronal heating in the energy equation, and we examine the resulting energy release processes in the vicinity of the flare and current sheet.  The numerical model is described in Section 2.  An overview of the simulated eruption is given in Section 3. The results of the energy partition analysis is given in Section 4, and simulated EUV and X-ray emission images are given in Section 5.  Discussion and conclusions are presented in Section 6.  

\section{Model description }\label{model.sec}

For this simulation, we use the coronal MHD code MAS developed by 
Predictive Science Inc. (see Appendix A in \citealt{Torok2018} and references therein), considering the same configuration as Model v1 in \citet{Mikic2013}.  The code solves the following MHD equations in spherical coordinates:
\begin{figure*}
\includegraphics[scale=0.3]{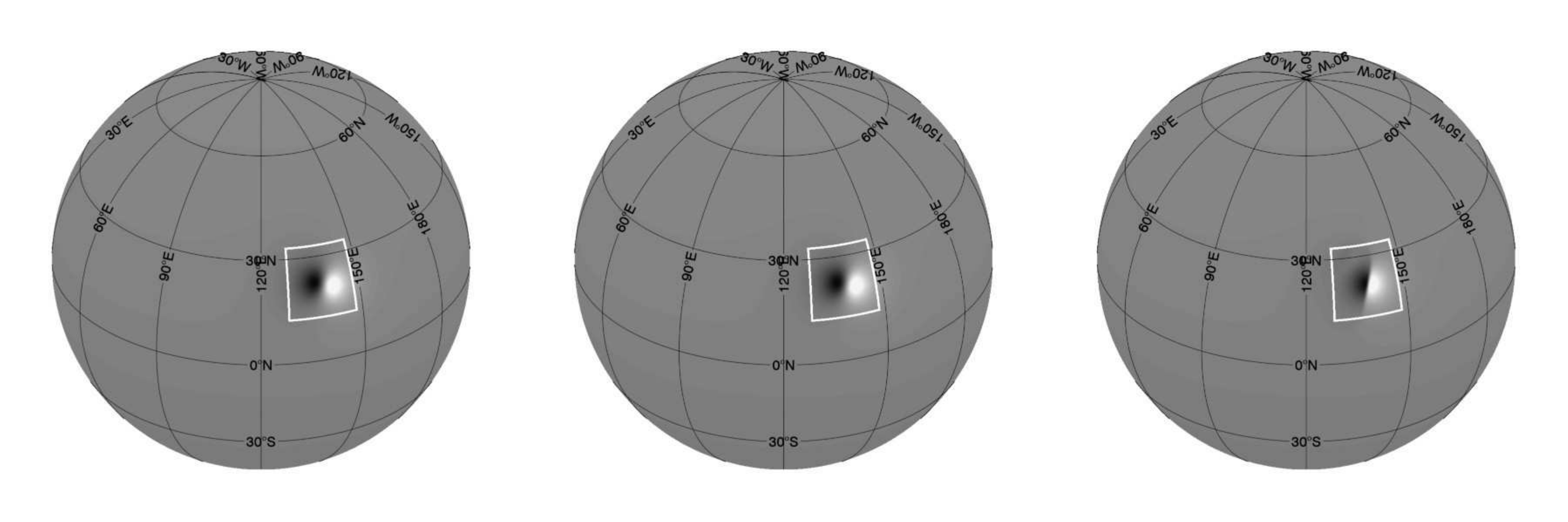}
\includegraphics[scale=0.4]{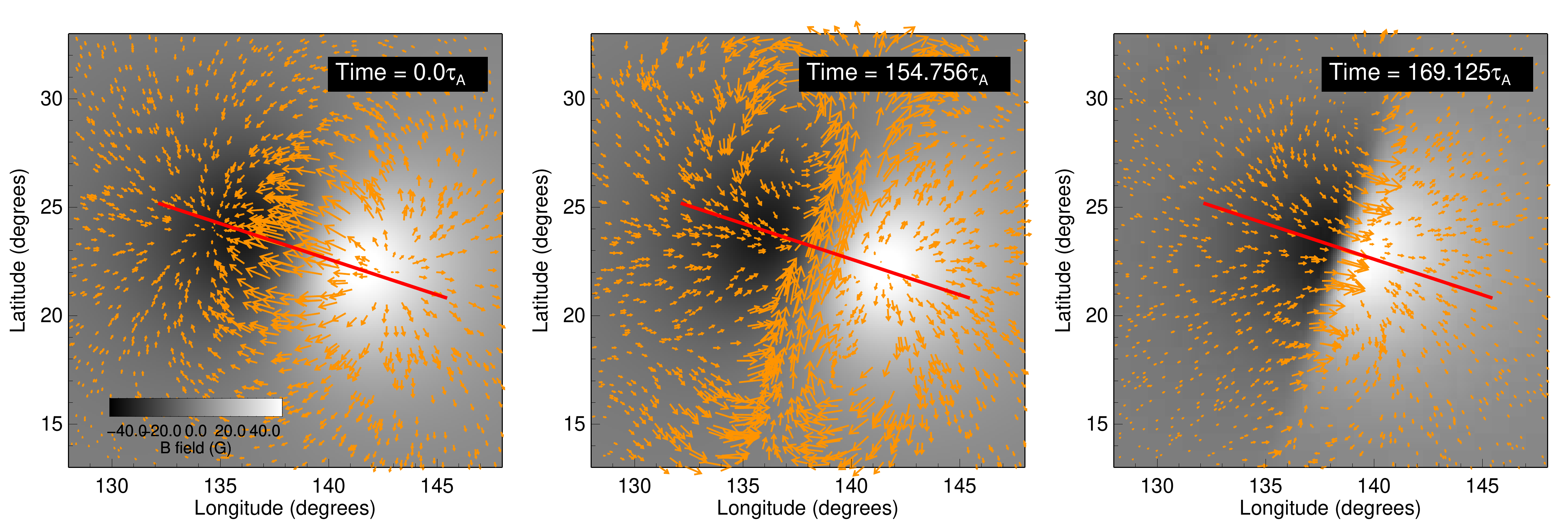}
\caption{\label{Br_boundary.fig}  The radial magnetic field ($B_r$) distribution at the lower boundary of the simulation (1 $R_s$), shown in the black and white color table for the full Sun (top row) and a close up of the portion of the Sun indicated by the white box (bottom row). Orange arrows on the bottom row of plots show the transverse field $\mathbf{B}_t= ({B}_\theta, {B}_\phi)$.  The left column shows the field 
at the beginning of the simulation (to show the initial magnetic field and boundary condition), 
the middle column shows the field after the transverse electric field emergence and subsequent relaxation, and the right column shows the field just before the eruption begins. The red line in the bottom row of plots indicates the location of the cut plane used in Section 3.}
\end{figure*}

\begin{eqnarray}
\nabla\times{\mathbf B} &=&  {4\pi\over c}{\mathbf J},\label{curlb-eq} \\
{1\over c}{\npar{\mathbf B}\over\npar t}  &=&
 -\mathbf\nabla\times{\mathbf E}, \label{faraday-eq} \\
{\mathbf E} + {1\over c}{\mathbf v}\times{\mathbf B} &=&  \eta {\mathbf J},
\label{ohm-eq} \\
\frac{\partial \rho}{\partial t} +\mathbf{ \nabla \cdot}(\rho \mathbf{ v})
&=& 0, \label{rho-eq} \\
\rho \left (\frac{\partial \mathbf{ v}}{\partial t}
 + \mathbf{ v \cdot}\mathbf{ \nabla} \mathbf{ v} \right ) &=&
 \frac{1 }{c} \mathbf{J}\times \mathbf{B}- \mathbf{ \nabla } p   \cr
  &+& \rho\mathbf{g} 
 + \mathbf{\nabla }\cdot (\nu \rho \mathbf{ \nabla v}), \label{mo-eq}
\\
\frac{1}{\gamma -1} \left (
\frac{\partial T}{\partial t}+ {\mathbf v} \cdot  \nabla T \right)
  & =& -~ T\nabla \cdot {\mathbf v}  + \frac{m}{2\rho k_B}S. \label{en-eq} 
\end{eqnarray}
In the above equations,  $\mathbf{B}$ is the magnetic field, $\mathbf{J}$ is the electric
current density, $\mathbf{E}$ is the electric field,
$\rho$, $\mathbf{v}$, $p$, and $T$ are the plasma mass density, velocity,
pressure, and temperature, respectively, 
${\mathbf g}=-g_0{\mathbf{\hat r}}{R_s}^2/r^2$
is the gravitational acceleration (where $R_s$ is the solar radius), $\eta$ is the resistivity, $\nu$ is the kinematic viscosity, $\gamma=5/3$ is the ratio of specific heats, $k_B$ is Boltzmann's constant, $m$ is the proton mass, and $S$ contains terms due to conduction, radiation, ohmic heating, and coronal heating.  The resistivity $\eta$ is uniform throughout the simulation volume, and is set via the Lundquist number ($S_L$) which has a value of $10^6$. The Lundquist number is defined as $S_L = \tau_R/\tau_A = 4\pi R_{s}^{2}/\tau_A\eta c^2$.  A typical Alfv\'{e}n speed at the base of the model corona is $V_A = 480$ km s$^{-1}$  (corresponding to $|B| = 2.2$ G and $n_0 = 10^8$ cm$^{-3}$), giving an Alfv\'{e}n crossing time ($\tau_A$) of 1446 s for a distance of 1 $R_s$ and a resistive diffusion time of $\tau_R = 4 \times 10^5$ hr.  

The terms in $S$ in Equation (\ref{en-eq}) are given by
\begin{equation}  
S = 
{-\mathbf\nabla}\cdot{\mathbf q} -n^2 Q_{rad}(T) + H_{\eta} 
+H_{ch} \label{S.eq}
 \end{equation}
where $-\nabla\cdot{\mathbf q}$ describes the heat transport due to thermal conduction, $-n^2 Q_{rad}(T)$ describes losses due to radiation, and $H_{\eta}$ and
$H_{ch}$ are heating terms due to ohmic dissipation and  coronal heating, respectively.  The thermal conduction term is collisional (i.e., Spitzer) in the lower corona, and collisionless  \citep[see][]{Hollweg1978}  higher up \citep{Lionello2001}, as prescribed by the following equations:
\begin{equation}
{\bf q} =  \left\{ 
\begin{array}{r@{\quad \quad}l}
-{\bf\kappa_0}T^{5/2}\mathbf{\hat{b}\hat{b}}\cdot\nabla T & r \lesssim 10R_{sun} \\ 
\frac{1}{(\gamma-1)}n_ekT{\bf v} & r \gtrsim 10R_{sun}
\end{array} \right.
\label{conduction.eq}
\end{equation}
where $\kappa_0 = 9 \times 10^{-7}$ erg K$^{-7/2}$ cm$^{-1}$ s$^{-1}$ and $\mathbf{\hat{b}}$ is the unit vector along the magnetic field.  The function makes a smooth transition between the two forms of conduction,  with the contribution from the collisional conductivity varying as $0.5(1-\tanh[0.2r-2])$ and the contribution from the collisionless conductivity varying as $0.5(1+\tanh[0.2r-2])$, where $r$ is in solar radii.

\begin{figure*}
\includegraphics[scale=0.12]{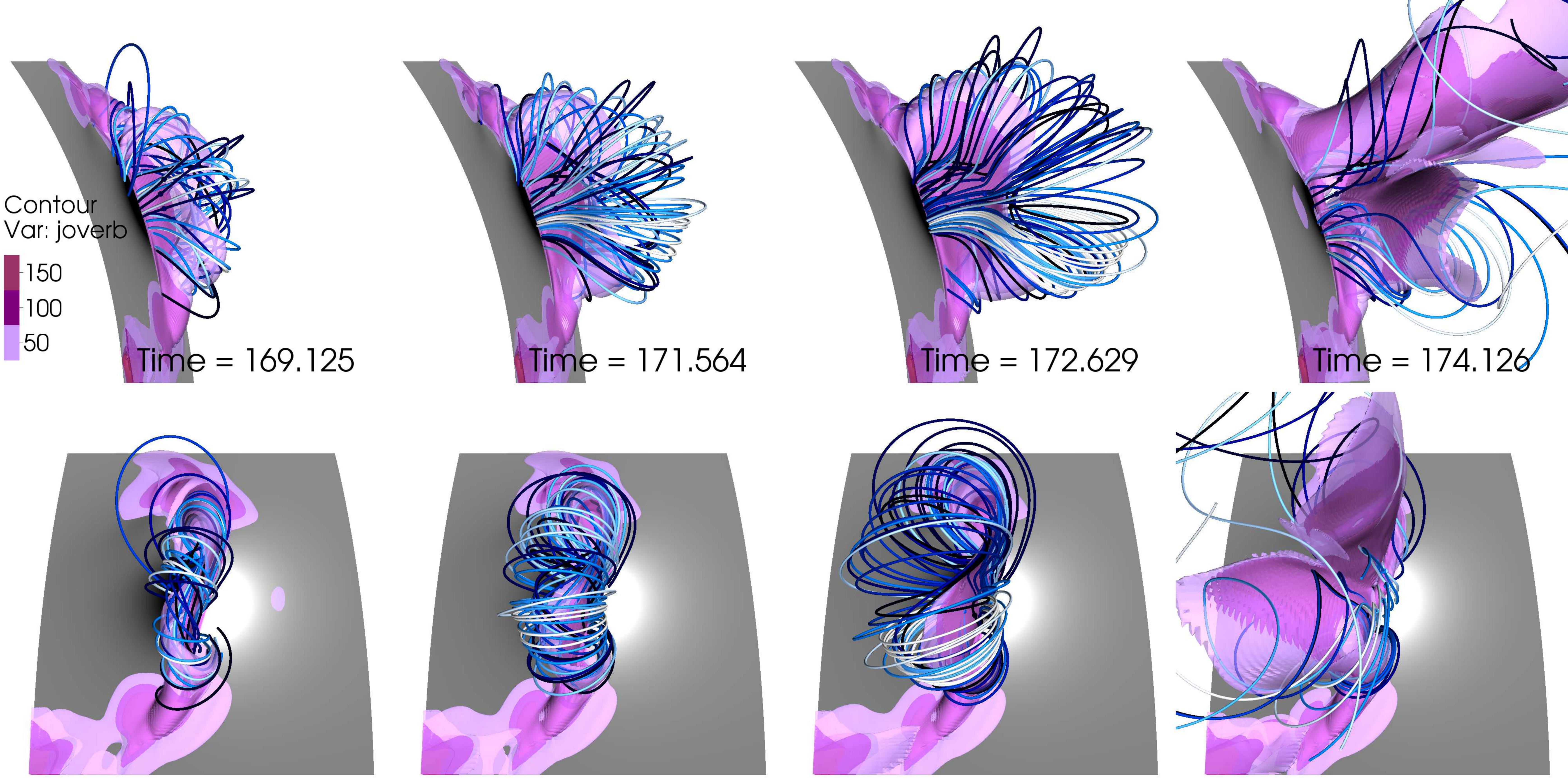}
\caption{\label{flux_rope.fig} Field lines (white/blue/black) showing the flux rope, and isovolumes of $|J|/|B|$ (pink/purple) indicating the location of the current sheet at several different times during the eruption.  Top row: view looking nearly perpendicular to the axis of the flux rope.  Bottom row: view from the top.  The radial B field at the surface is also shown in black and white. $|J|/|B|$ isovolume contours are given in normalized code units. (An animation of this figure is available.)}
\end{figure*}

In Equation (\ref{S.eq}),  $Q_{rad}(T)$ is a standard radiative loss function that has a maximum near $10^5$ K \citep[see][]{Athay1986},  and $n$ is the electron number density (which is taken to be the same as the proton number density in the case of a hydrogen plasma).  The ohmic heating term is given by $H_\eta = \eta J^2$.   The mechanism that heats the corona is currently unknown, so for the coronal heating term $H_{ch}$, we use an empirical three-part function containing an exponential heating term, a term designed to simulate heating along polarity inversion lines in the quiet sun, and an active region heating term. The form of this heating equation is given in \citet{Lionello2009}, and that work showed that this three-part heating function gives reasonable results when comparing to observed emission from EUV and X-ray telescopes.

The techniques used to solve Equations (\ref{curlb-eq})--(\ref{S.eq}) have been discussed in detail elsewhere \citep{Mikic1994,LinkerMikic1997,Lionello1999,Mikic1999,Linker2001, Lionello2001,Linker2003}.   We note that solutions to these equations include a numerical technique to artificially broaden the transition region without affecting the coronal part of the solution in order to avoid strong gradients in the transition region \citep{Lionello2009,Reeves2010}.   In order to achieve this broadening effect in the transition region, $\kappa$ is increased and $Q(T)$ is decreased at low temperatures such that that the product $\kappa(T)Q(T)$ is unchanged.   In our simulation, for temperatures below $T_c$ = 500,000 K,  $\kappa(T)$ is multiplied by a factor of $(T_c/T)^{2.5}$ and $Q(T)$ is divided by that same factor so that $\kappa(T)Q(T)$ remains constant. 

The boundary condition at 1 R$_s$ is a magnetic flux distribution produced by a localized subsurface dipole superimposed on a global dipole field to mimic the presence of an active region within the global corona.  The latitude and orientation of the dipole are similar to those of NOAA active region number 8038 which produced a CME and an associated C1.3 class solar flare on 12 May 1997.   The observations of this event have been well-studied \citep[e.g.][]{Thompson1998,Webb2000,Attrill2006} and several different groups have modeled it \citep[e.g.][]{Wu2007,Cohen2008,Titov2008}.  We use a more simplified magnetic field distribution at 1 R$_s$ than in previous models of the 12 May 1997 event, in order to investigate a configuration based on a realistic case without introducing unnecessary complexity. The left column of Figure \ref{Br_boundary.fig} shows the global radial magnetic field at the photosphere in the top row along with a close-up showing detail in the active region in the bottom row.  In this figure, the black and white color table represents the radial magnetic field ($B_r$) at the lower boundary and the orange arrows indicate the transverse magnetic field $\mathbf{B}_t = (B_\theta, B_\phi)$.  The upper radial boundary is at 20 R$_s$, where the flow is supersonic and super-Alfv{\'e}nic, and the boundary conditions are such that only outgoing waves are permitted, allowing plasma and magnetic fields to flow freely out of the simulation domain.  A full accounting of the boundary conditions and their implementation can be found in \citet{LinkerMikic1997,Mikic1999,Linker2001}.

The equations are solved on a non-uniform, spherical mesh of size $356\times351\times261$ ($r$,$\theta$,$\phi$) mesh points, with the smallest grid scales located in areas of interest. The smallest mesh cells in the $r$ direction are $\sim$63.6 km, and they are located near the lower boundary, from $r=1R_s$ to $r=1.0146R_s$.   The smallest mesh cells in $\theta$ and $\phi$ are located in the active region.  The smallest $\theta$ mesh cells cover $\sim$0.11665 degrees of latitude, and are located between 12.7241 degrees and 33.3715 degrees latitude.  The smallest $\phi$ mesh cells cover $\sim$0.11490 degrees of longitude, and are located between 135.300 degrees and 142.194 degrees longitude.

The initial conditions for the simulation are as follows.  The initial magnetic field is a potential field derived from the $B_r$ distribution shown in the upper left panel of Figure \ref{Br_boundary.fig}.  The initial temperature, density, and velocity of the plasma are given by a spherically symmetric solar wind solution for the specified heating.   The MHD equations are integrated in time until a quasi-steady MHD solution is reached at $t =122.5\tau_A$ (i.e., 49.2 hrs).  An idealized solar minimum configuration is formed, with a streamer belt and polar coronal holes.  We refer to this phase of the simulation as the relaxation phase.

\section{Overview of Simulated Eruption \label{overview.sec}}

The goal of this simulation is to store energy in the magnetic field and then trigger its rapid release, causing an eruption.  After the initial relaxation, the magnetic field in the active region is still close to a potential state. There are several methods for achieving energization of the field, including imposing a converging or shear flow at the active region \citep[e.g.][]{Mikic1994,Reeves2010,Karpen2012},  driving a flux rope through the lower boundary \citep[e.g.][]{Fan2003,Fan2007,Fan2011,Fan2016,Dacie2018}, or inserting a flux rope into a pre-existing 
background field \citep[e.g.][]{Aad2004,SavchevaPariat2012,Torok2018}.  The eruption can be triggered in a variety of ways, including an ideal MHD instability such as the kink instability \citep[e.g.][]{TorokKliem2005}, or a loss of equilibrium brought on by boundary flows \citep[e.g.][]{LinForbes2000,ReevesForbes2005, Reeves2007}, flux emergence \citep{Chen2000,Lin2001}, or cancelling flux at the active region \citep[e.g.][]{Reeves2010,Amari2010}.  An extensive compilation of eruption trigger mechanisms can be found in \cite{Green2018}.

In our simulation, we energize the fields by introducing a transverse magnetic field, $\mathbf{B}_t$, along the polarity inversion line in the active region.  This process is accomplished by applying a transverse electric field $\mathbf{E}_t = \nabla\Phi_t$ at the boundary $r = R_s$.  We specify $\Phi$ such that the injected $\mathbf{B}_t$ is aligned along the polarity inversion line, causing stress in the fields. The transverse field ``emergence'' begins at $t =122.5\tau_A$ and continues until $t=152.5\tau_A$. The field is then allowed to relax until $t=155\tau_A$.  The middle column 
of Figure \ref{Br_boundary.fig} shows $B_r$ (black and white image) and $\mathbf{B}_t$ (orange arrows) at the end of this relaxation phase.  The orange arrows outline the sheared, non-potential fields that have been introduced to the active region. The addition of $\mathbf{B}_t$ does not modify $B_r$ at the photosphere, as can be seen by comparing $B_r$ in the left and middle panels of Figure \ref{Br_boundary.fig}.  At the end of the relaxation phase at $t=155\tau_A$, there is not much twist in the newly emerged non-potential field, except for a small amount that is created by reconnection between the emerging flux and the pre-existing potential field.

At $t=155\tau_A$, flux cancellation is initiated by imposing converging flows, together with photospheric diffusion \citep[e.g.][]{Amari2010,Bisi2010,Mikic2013}.  As it progresses, the flux cancellation adds the majority of the twist to the emerged fields, forming the flux rope \citep[e.g.][]{Aad1989}.  The right column of Figure \ref{Br_boundary.fig} shows the magnetic field configuration at the surface after the flux rope has formed and just before the eruption, at $t=169.125\tau_A$.  The orange arrows in the bottom right panel of Figure \ref{Br_boundary.fig} are pointed opposite from the potential field configuration shown in the bottom left panel.  This configuration is the signature of a ``bald patch'' \citep{Titov1993} magnetic field configuration, which is sometimes seen in observations \citep[e.g.][]{LopezAriste2006,LiuL2019}, and is indicative of the formation of a coronal flux rope above the polarity inversion line.

Figure \ref{flux_rope.fig} shows magnetic field lines and isovolume surfaces of $|J|/|B|$ for the simulation just before and during the eruption.  Just before the eruption occurs, at $t=169.125\tau_A$, the magnetic field lines outlining the flux rope have taken on an inverse-S shape, and the current sheet also has this shape, which is common in 3D simulations of flux ropes \citep[e.g.][]{Fan2004,Gibson2006} and sigmoidal active regions \citep[e.g.][]{Savcheva2014}.  As the flux cancellation progresses, at around at $t=170\tau_A$, the flux rope loses stability and begins to erupt.  During the main part of the eruption (e.g. at $t=172.629\tau_A$ in Figure \ref{flux_rope.fig}) the strongest parts of the current sheet are in the center of the active region, where the overlying field lines are stretched the most.  At about $t=173.5\tau_A$, when the flux rope has already reached a significant height, the current sheet structure begins to rotate clockwise as viewed from above. This rotation can be seen in the lower right panel of Figure \ref{flux_rope.fig} at $t=174.126\tau_A$, where the southern part of the current sheet structure has moved to the left with respect to the previous panel, and the northern part of the current sheet has moved to the right (see also the movie that accompanies Figure \ref{flux_rope.fig}).  The current sheet is quite curved at this point, wrapping around the legs of the erupting CME.

We note that, since our flux rope has left-handed twist (negative helicity), the direction of the rotation is opposite to what is expected from the conversion of flux-rope twist into writhe \citep[e.g.,][]{Green2007,Torok2010}. Indeed, rotation due to twist conversion should occur right from the onset of the eruption, which is not the case here, indicating that this effect does not play a role in our eruption. Rather, the rotation may be caused by the interaction of the flux rope with the large-scale dipole field, once the rope has ascended to a height at which this field becomes significant. The large-scale dipole field runs from north to south in our model and therefore constitutes an external shear field for the flux rope, the presence of which can induce rotation \citep{Isenberg2007,Kliem2012}. We will leave this question, which is not relevant for the analysis presented here, to a later investigation.

\begin{figure*}
\includegraphics[scale=0.6]{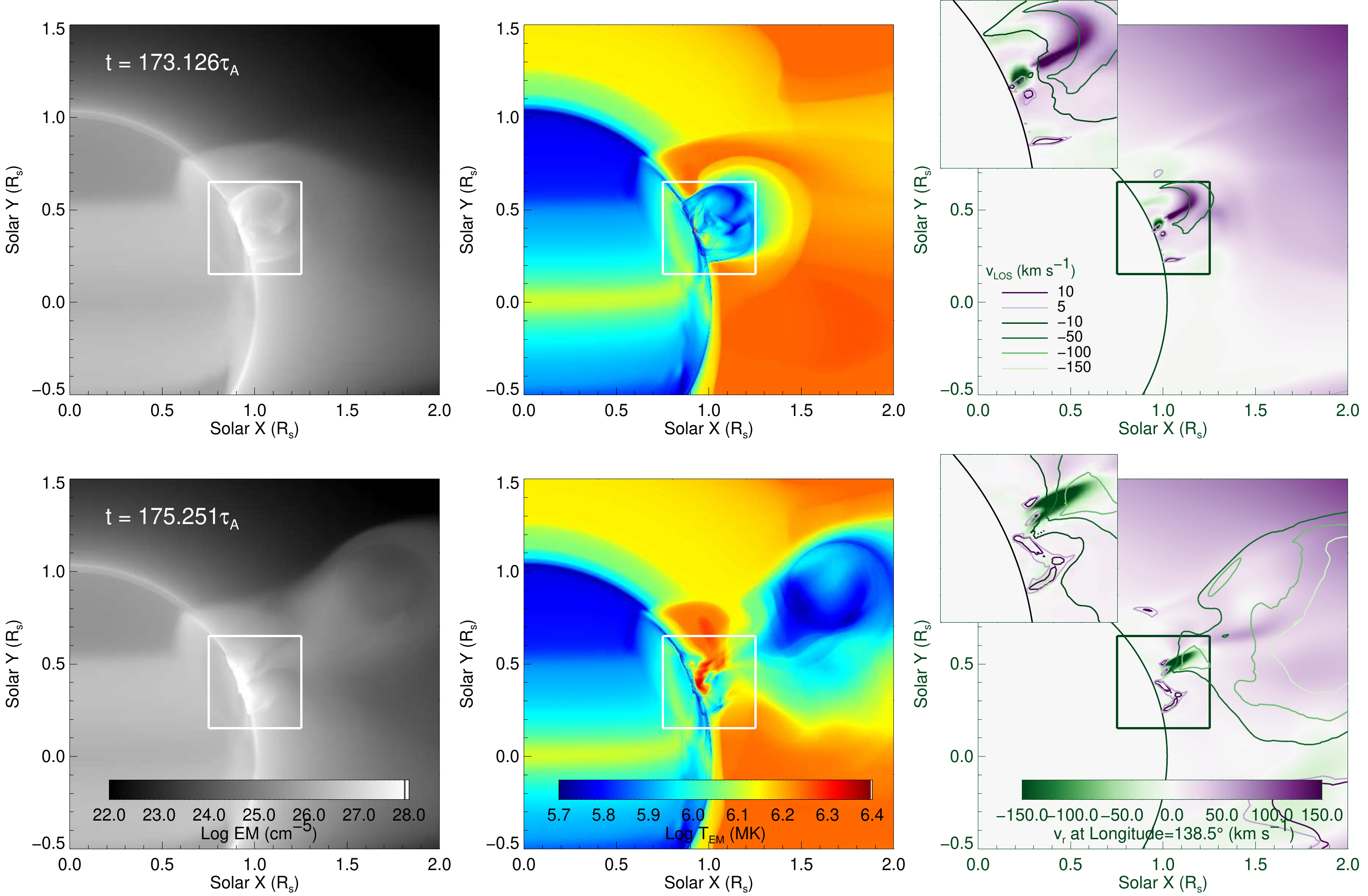}
\caption{\label{cutplanes_big.fig}  
Log of the emission measure (EM, left), 
log of the EM-weighted temperature ($T_{EM}$, middle), and a radial velocity ($v_r$) slice in a plane at a longitude of 138.5 degrees (right) for two times during the eruption. The $v_r$ plot also shows contours for the 
EM-weighted line-of-sight velocity ($v_{LOS}$) .  For the $v_{LOS}$ contours, green contours indicate velocities towards the observer.  The box indicates the field of view of the inset in the $v_{r}$ panels, as well as the field of view for Figures \ref{volume_side.fig}-\ref{joverb_emission.fig}.}
\end{figure*}

Figure \ref{cutplanes_big.fig} shows the emission measure (EM), EM-weighted average temperature ($T_{EM}$), and a slice of the radial velocity, $v_r$, for two times during the eruption.  The emission measure is given by
\begin{equation}
EM = \int n^2dl
\end{equation}
where $n$ is the number density of the plasma and $l$ is the line of sight distance.  $T_{EM}$ is given by
\begin{equation}
T_{EM}   =  \frac{\int Tn^2dl}{EM} 
\end{equation}
where $T$ is temperature.  The $v_r$ slice is taken at a longitude of 138.5 degrees, which cuts through the center of the magnetic bipole, and insets in the $v_{r}$ panels show a close up of the reconnection region in order to show more detail there.  The $v_r$ plots also show contours of the EM-weighted line-of-sight velocity ($v_{LOS}$) in order to indicate motion out of the plane of the image, which is given by
\begin{equation}
v_{LOS}  =  \frac{\int \mathbf{v}\cdot\mathbf{\hat{\ell}}n^2dl}{EM}
\end{equation}
where $\mathbf{v}$ is the velocity vector and $\hat{\ell}$ is a unit vector in the direction of the line of sight.  In the $v_r$ plots, purple (positive) indicates velocity radially away from the Sun, and green (negative) indicates velocity radially towards the Sun.  For the $v_{LOS}$ contours, purple (positive) indicates velocity into the page (away from the observer), and green (negative) indicates velocity out of the page (towards the observer).

The flux rope is visible in both the EM and $T_{EM}$ plots in Figure \ref{cutplanes_big.fig}, and the effects of the aforementioned clockwise rotation can be seen by comparing the two times.  At $t=173.126\tau_A$ the flux rope is dense and cool, so it shows up as a bright structure in the EM plot and 
as a blue structure in the $T_{EM}$ plot, indicating an EM weighted average temperature of 0.6 - 1 MK.  At this time the flux rope axis is approximately parallel to the image plane.  At $t=175.251\tau_A$, the flux rope has expanded and rotated so that the axis is more along the line of sight, and the flux rope appears as a large, somewhat faint tear-drop shaped structure in the EM plot.  The EM weighted average temperature of the flux rope is about 0.6 MK, as seen in the $T_{EM}$ plot.  The $v_{LOS}$ contours in the right panels show that the motion of the flux rope during the eruption is angled towards the observer, resulting in a line of sight velocity on the order 50 km s$^{-1}$ at the first time, and 150 km s$^{-1}$ later in the eruption.  The flux rope has a radial velocity of about 100 km s$^{-1}$ at the earlier time, and a radial velocity of 200-250 km s$^{-1}$ at the later time.  These high radial velocities are not visible in the lower right panel of Figure \ref{cutplanes_big.fig} because the CME has rotated out of the plane at 138.5 degrees longitude.

Underneath the flux rope, magnetic fields reconnect and form the flare loops. These reconnected loops show up as features with high EM values low in the corona in Figure \ref{cutplanes_big.fig}. At the earlier time in Figure \ref{cutplanes_big.fig}, the temperatures in this region are still relatively cool, though there is some hotter ($\sim$3 MK) plasma low down that is difficult to see in this global view (see the discussion in later sections for more details).  At the later time, there is a prominent area of plasma with temperatures $\ge$2 MK low in the corona, indicating that significant and widespread heating has occurred.  Opposing flows, as would be expected in a reconnection outflow, are seen in the insets for the $v_r$ images at both times, with outflow speeds up to 200 km s$^{-1}$.  The $v_{LOS}$ contours at the earlier time also show opposing flows along the line of sight low in the corona, with much smaller velocities in the range of 5-10 km s$^{-1}$.  This velocity indicates that the reconnecting structure is tipped slightly towards the observer, so that there is a small component of the reconnection outflow in the line of sight direction as well.  The $v_r$ inset at the later time shows loop-like features near the Sun, with small downward velocities of 5-10 km s$^{-1}$, and the $v_{LOS}$ contours at this time show a loop-like structure with a small velocity of 5-10 km s$^{-1}$ away from the observer (purple contours).  These features indicate newly reconnected loops that are tilted towards the observer and are shrinking and becoming more potential, as has been seen in observations \citep[e.g.][]{ForbesActon1996,Reeves2008}.

\begin{figure}
\includegraphics[bb=20 0 1000 450,clip=true,scale=0.45]{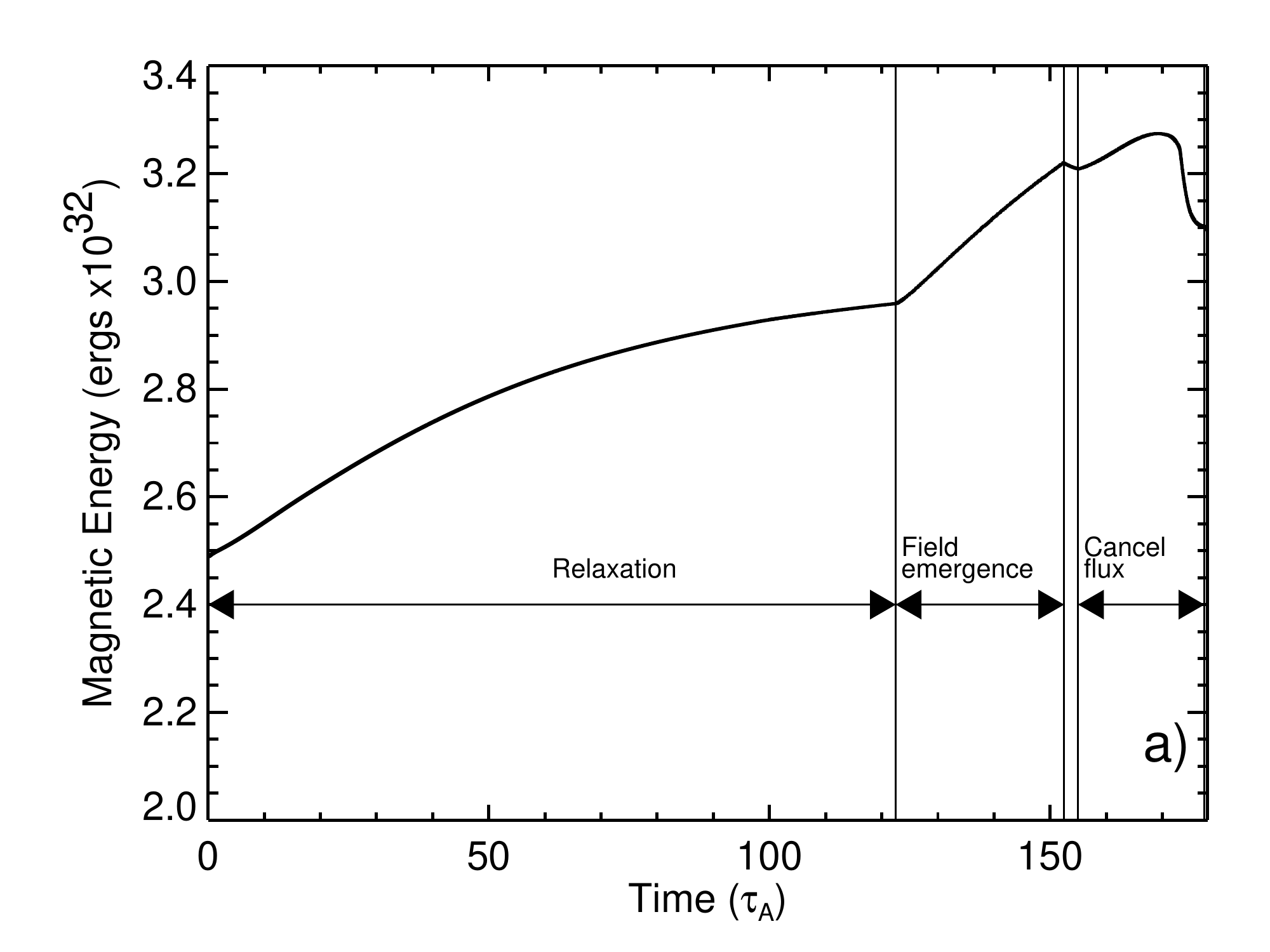}
\includegraphics[bb=20 0 1000 450,clip=true,scale=0.45]{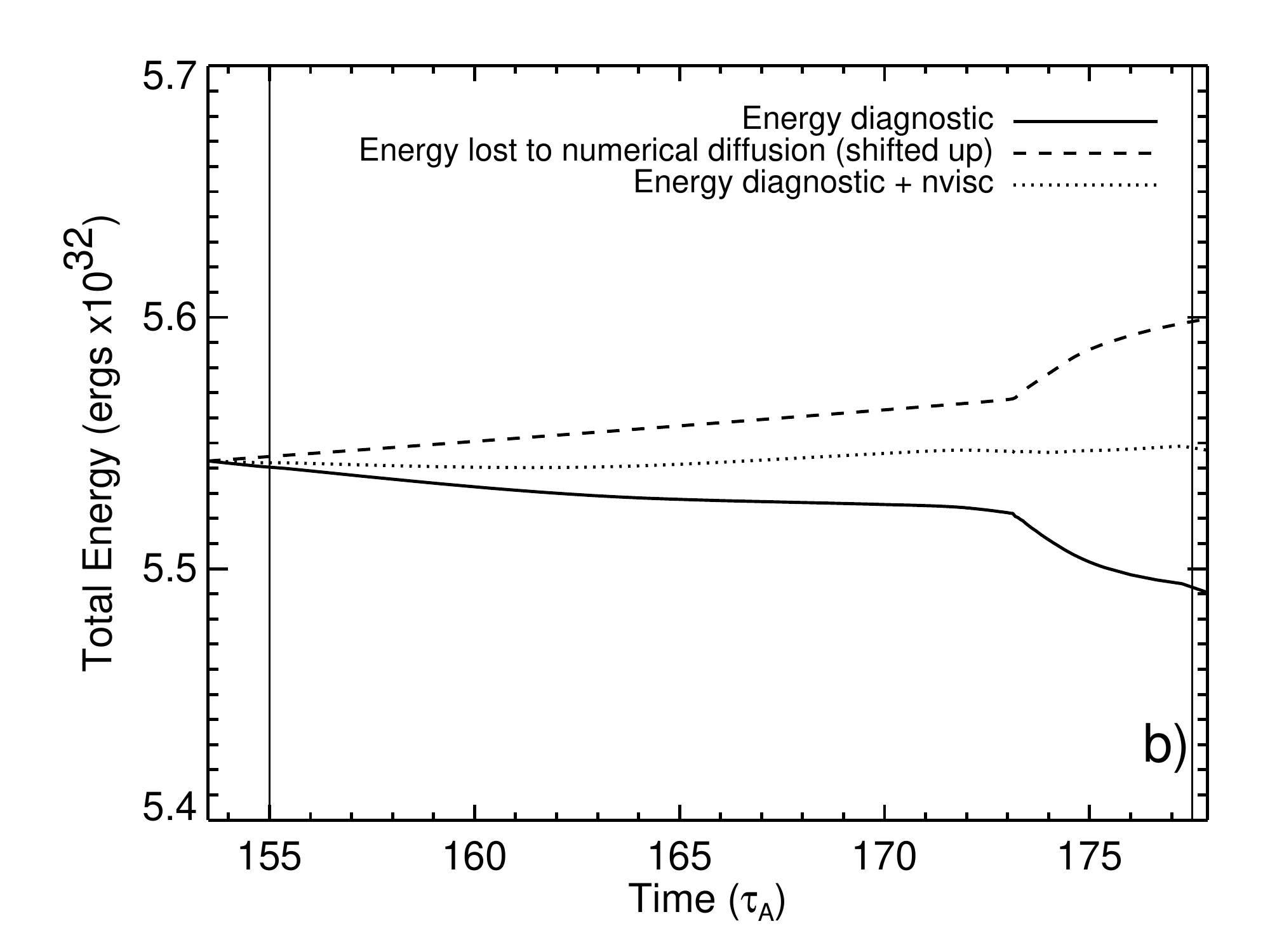}
\caption{\label{global_energy.fig}  Panel a) shows the evolution of the total magnetic energy over the entire simulation domain, with different phases of the simulation demarcated by vertical lines.  Panel b) shows the energy conservation diagnostic (solid line), the heating due to numerical dissipation in the code (dashed line, shifted up by 5.54$\times 10^{32}$ ergs for display purposes), and the sum of these two terms (dotted line).  Vertical lines show the beginning and ending of the flux cancellation phase. }
\end{figure}

\section{Energy Diagnostics \label{results.sec}}

The global magnetic energy integrated over the simulation volume is plotted in Figure \ref{global_energy.fig}a.  The energy increases from 2.49$\times$10$^{32}$ ergs to 2.96$\times$10$^{32}$ ergs during the relaxation phase due to the influence of the solar wind, which tends to open previously closed field lines.  The magnetic energy increases more steeply during the time period where transverse field is 
introduced, up to 3.22$\times$10$^{32}$ ergs.  After the end of that phase, the energy decreases slightly as the simulation is allowed to relax.  During the flux cancellation phase, the energy increases to a maximum of 3.27$\times$10$^{32}$ ergs at about $t=170\tau_A$, when the eruption begins, and decreases to 3.09$\times$10$^{32}$ ergs during the eruption.  

A global energy diagnostic, similar to the one developed for the 2D simulation in \citet{Reeves2010}, is shown in Figure \ref{global_energy.fig}b.  This diagnostic is the sum of the global magnetic, internal, and kinetic energies, viscous dissipation, work done against gravity, changes in energy due to radiation and heat sources, and the energy deposited in or carried away from the simulation domain due to flows, e.g., conductive flux, kinetic energy flux, enthalpy flux, and Poynting flux.  A constant value of this diagnostic as a function of time indicates that energy is conserved in the simulation.  Figure \ref{global_energy.fig}b shows that the calculated total energy after the eruption is  $\sim$0.056 $\times 10^{32}$ ergs less than the value at the end of the field emergence phase, so some energy is being lost due to numerical dissipation.  

We estimate the coefficient of numerical diffusion in the code due to ``upwinding'' with the following equation:
\begin{equation}
\nu_N = \sum_{i=r,\theta,\phi}\frac{v_i\Delta x_i}{2}\left(1 - v_i\Delta t/\Delta x_i\right)
\end{equation}
where $\Delta x_i$ is the grid spacing for the $i$th dimension, i.e. $(\Delta r, r\Delta\theta, r\sin\theta\Delta\phi)$, $v_i$ is the plasma velocity in the cell in the $i$th dimension, and $\Delta t$ is the time step in the code.  Because of the variable size mesh, the magnitude of $\nu_N$ varies over the simulation domain, but in the region we are interested in, it is on the order of $10^{-4}$ or less. We calculate the excess viscous heat produced by the numerical diffusion using the equation
\begin{equation}
H_{nvisc} = \rho\nu_N\left(\frac{1}{2}e_{ij}e_{ij} - \frac{2}{3}(\nabla\cdot\mathbf{v})^2\right)
\label{visc_heat.eq}
\end{equation}
where $e_{ij}$ is the rate of strain tensor. Integrating equation \ref{visc_heat.eq} over volume and time gives the amount of energy lost to numerical diffusion over the simulation volume. We plot this value in Figure \ref{global_energy.fig}b, shifted up by a constant value so that it is visible on the plot.  The sum of our energy diagnostic and the energy lost because of viscous heating due to the numerical diffusion is also plotted in Figure \ref{global_energy.fig}b. The sum is nearly constant over the flux cancellation phase, indicating that our estimate of the numerical losses in the code is quite accurate.

\subsection{Energetics in the current sheet}
The background magnetic field configuration is not perfectly symmetric in the simulation,  since the sub-photospheric dipole that was used to produce the active region is slightly tilted, and the region is located in the northern hemisphere, where the global background field is positive. This configuration leads to an eruption that is somewhat asymmetric and deflects to the east of the active region, as can be seen in Figure \ref{flux_rope.fig} and indicated by the $v_{LOS}$ contours in Figure \ref{cutplanes_big.fig}.  Nevertheless, it is instructive to examine various quantities in a plane that is perpendicular to the polarity inversion line that cuts through the center of the flux rope.  The location of the plane at 1 R$_s$ is shown superimposed on the radial magnetic field in Figure \ref{Br_boundary.fig}.  The plane extends from 1-1.2 R$_s$ in the radial direction.  The global plots in  Figure \ref{cutplanes_big.fig} show that the CME extends past the upper radial edge of this plane at 1.2 R$_s$, especially at later times, but here we will focus on the heating that occurs during the flare and in the lower part of the current sheet.

\begin{figure*}
\includegraphics[scale=0.82]{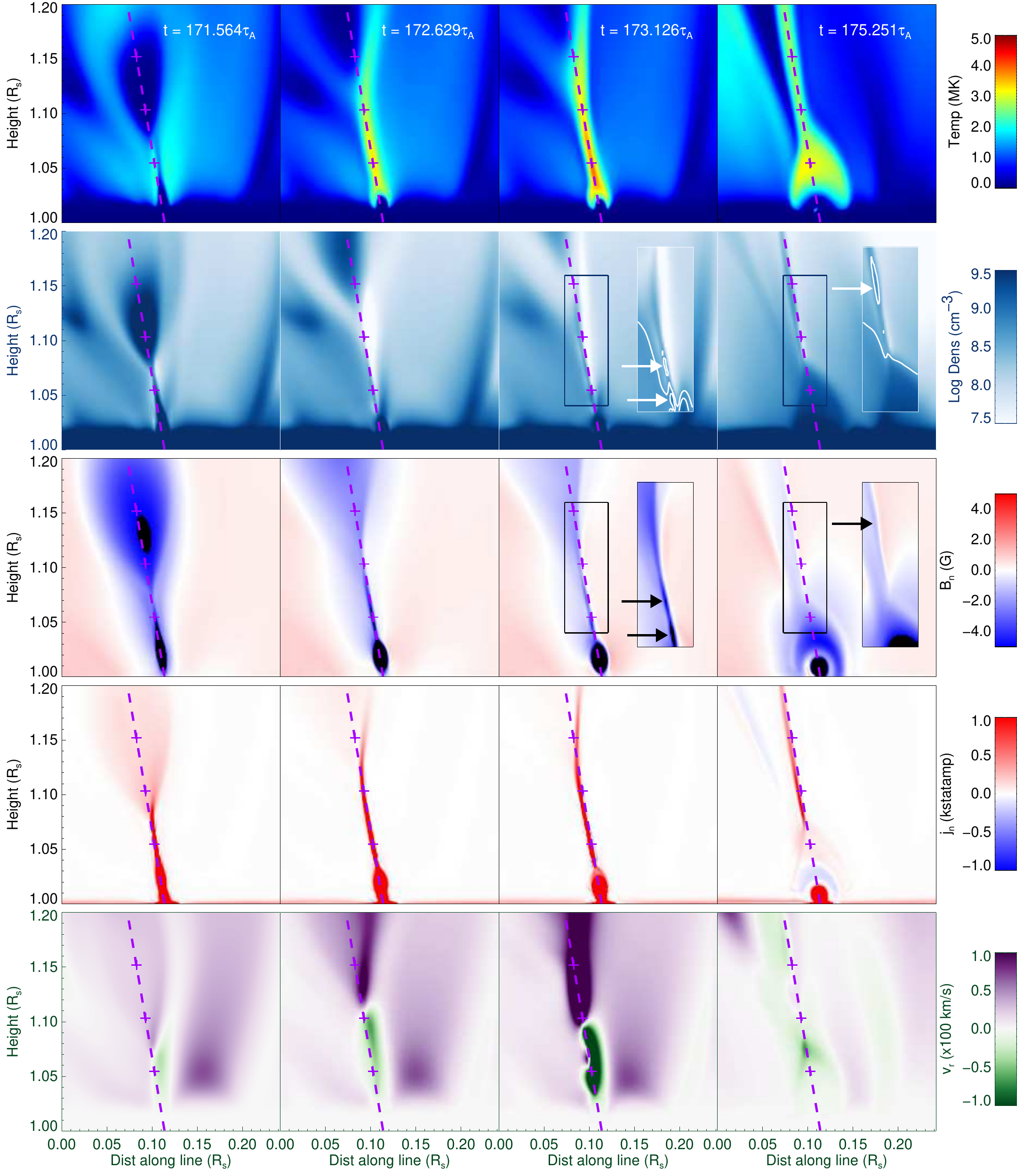}
\caption{\label{cutplanes_fields.fig} Temperature (top row), Log of density (second row), magnetic field normal to the plane (third row), current normal to the plane (fourth row) and radial velocity (bottom row) for the cut plane (red line) defined in Figure \ref{Br_boundary.fig}, shown at different times during the eruption. The dashed line indicates the line along which quantities are plotted in Figure \ref{heat_along_line.fig}, and crosses indicate distances of $r=1.05, 1.10, 1.15$ along the line.  The insets at the last two times in the density and B$_n$ images show details 
in the current sheet region, and arrows indicate positions of ``plasmoids.''  The inset for the density images includes contours that outline the plasmoids.}
\end{figure*}

Figure \ref{cutplanes_fields.fig} shows the evolution in the perpendicular plane of the temperature, density, radial velocity, and  the magnetic field and current normal to the plane.  This method of displaying the simulation results reveals features familiar from 2D simulations.  A cool and dense region of plasma forms the erupting flux rope, with a heated current sheet forming beneath it.  The evolution of these features is fairly smooth in the early stages of the eruption, but starting at about $t=173\tau_A$, the current sheet region starts to form dense plasmoid-like structures along its length.  These structures are also visible in the magnetic field normal to the cut plane, and they are close in size to the grid spacing of the simulation at the location of the current sheet.  Insets highlighting these structures are shown in the density and B$_n$ images in Figure \ref{cutplanes_fields.fig}, with arrows indicating their locations. It is worth noting that these structures are not strictly plasmoids as appear in 2D simulations, but rather rope-like regions of enhanced temperature and density that extend out of the page.  However, we will refer to them as ``plasmoids'' henceforth for the sake of brevity.

\begin{figure*}
\includegraphics[scale=0.82]{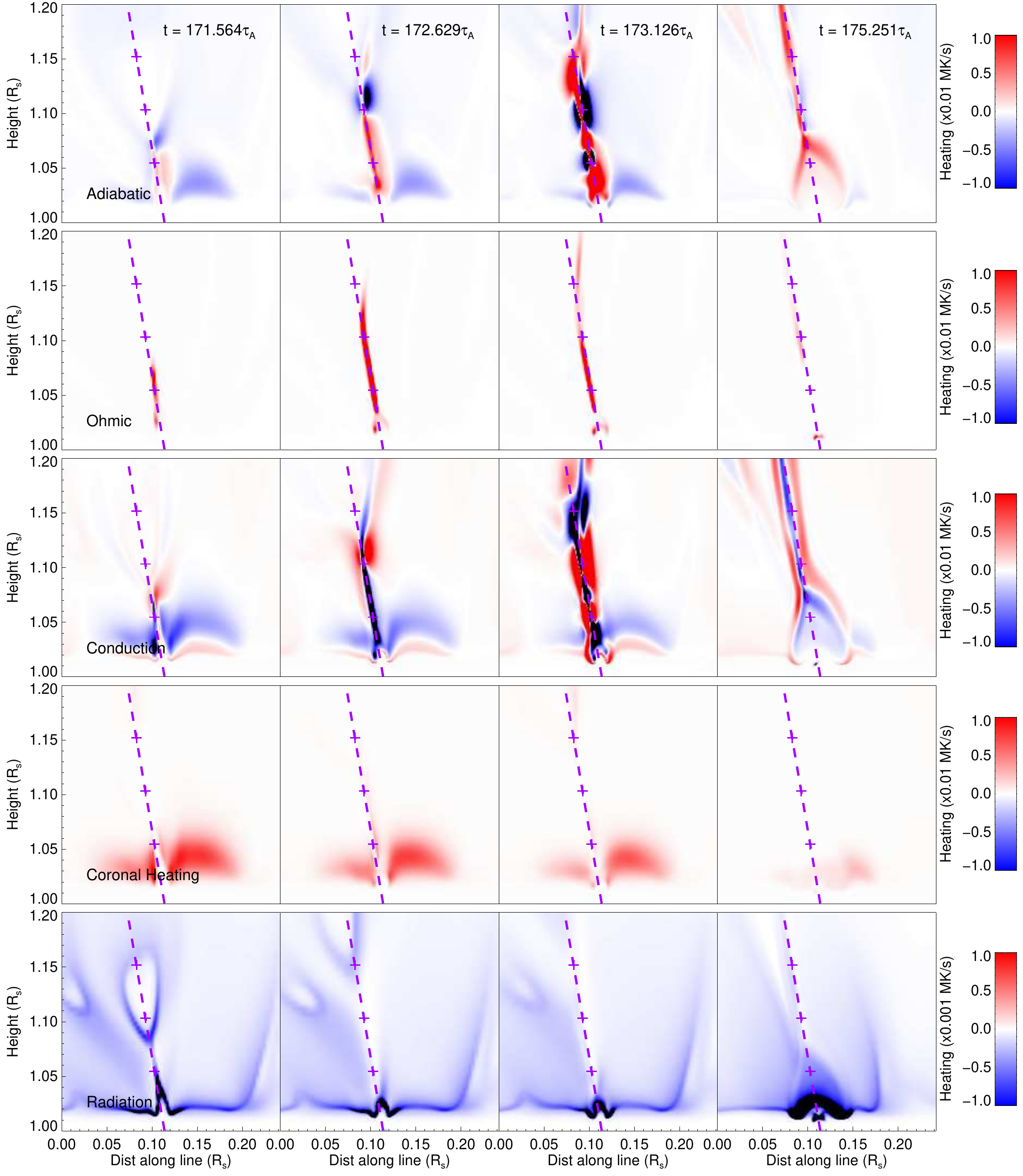}
\caption{\label{cutplanes_heat.fig} Terms that contribute to the heat generation, loss, and transfer in the plasma: adiabatic term (top row), contribution of ohmic heating (second row), thermal conduction (third row), coronal heating (fourth row),  and radiative cooling (bottom row) for the cut plane (red line) defined in Figure \ref{Br_boundary.fig}, shown at 
the same times as in Figure\,\ref{cutplanes_fields.fig}.  The scale for the radiation term is ten times smaller than the other terms so that features can be clearly seen.   The dashed line and crosses have the same meaning as in Figure \ref{cutplanes_fields.fig}.}
\end{figure*}

In order to understand how the plasma is heated in the current sheet region, we rewrite Equation (\ref{en-eq}) in the following way:

\begin{equation}
\frac{\partial T}{\partial t}= -{\mathbf v}\cdot \nabla T - (\gamma-1)T\nabla\cdot {\mathbf v} +\frac{m(\gamma -1)}{2\rho k_B} S \label{dTdt.eq}
\end{equation}  
The sources and sinks of the change in temperature in this equation are the adiabatic term, which is given by $-(\gamma-1)T\nabla\cdot {\mathbf v}$, and the ohmic heating, coronal heating, and radiation terms encompassed in $S$ (see Equation (\ref{S.eq})).  The terms responsible for transporting the temperature away from a certain location are the advection term, given by $-{\mathbf v}\cdot\nabla T$, and the thermal conduction term, $-\nabla\cdot{\mathbf q}$.  These heating terms are calculated in the cut plane indicated in Figure \ref{Br_boundary.fig} and the results are shown in Figure \ref{cutplanes_heat.fig}.

\begin{figure*}
\includegraphics[scale=0.64]{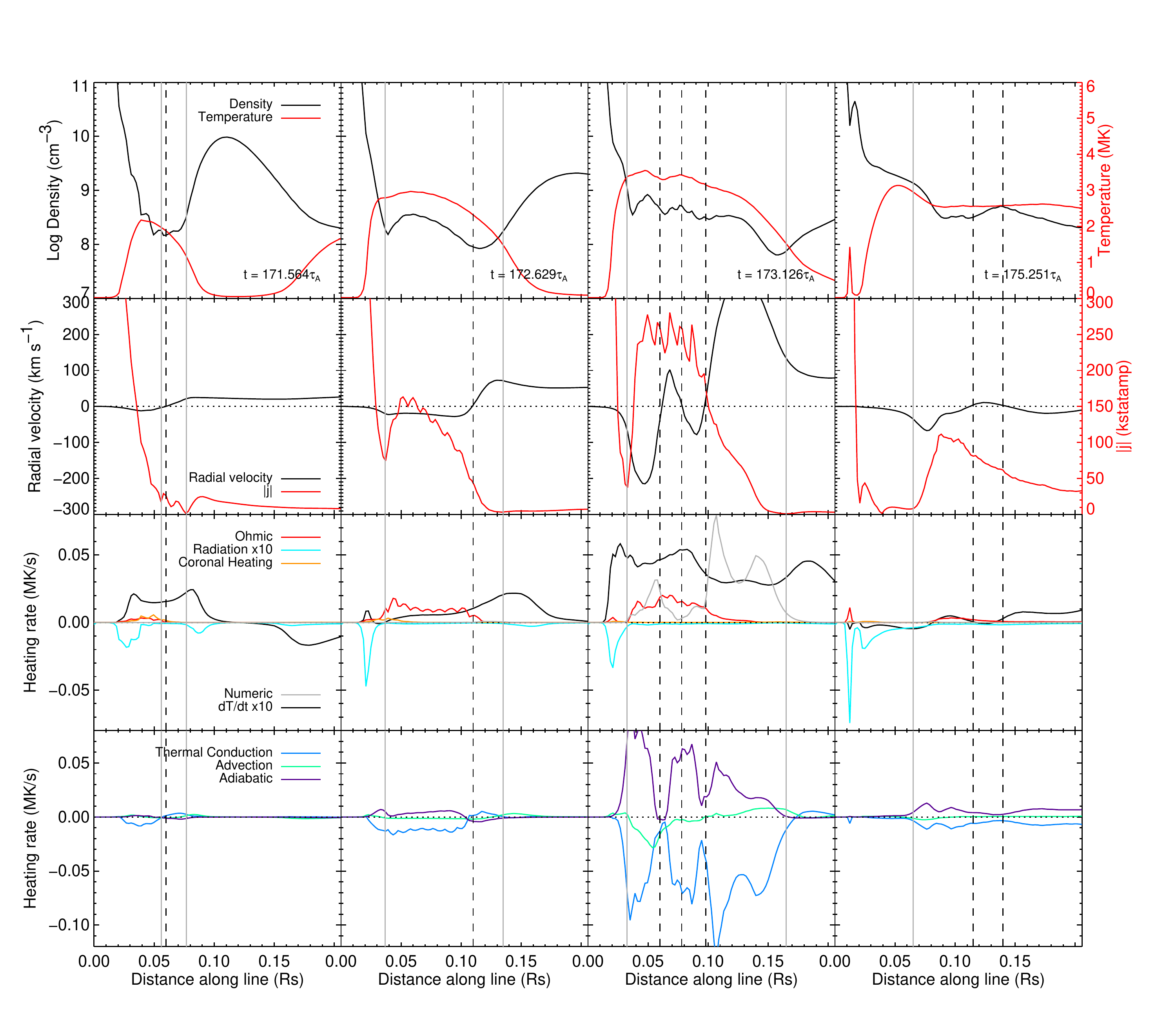}
\caption{\label{heat_along_line.fig} Plots of quantities along the dashed line in Figures \ref{cutplanes_fields.fig} and \ref{cutplanes_heat.fig}, shown at the same times as in those figures: temperature and density (top row), radial velocity and current ($|j|$) (second row), the change in the temperature ($dT/dt$) and source terms for heating and cooling (third row), and heat transport terms (bottom row). 
Stagnation points in the radial velocity are indicated by dashed vertical lines, and the current sheet ends are marked with grey vertical lines.}
\end{figure*}

Early in the eruption, the heating terms evolve fairly smoothly.  At $t=171.564\tau_A$ and $t=172.629\tau_A$, radiative cooling is elevated at the bottom of the current sheet, where the flare loops are.  There is also elevated radiative cooling around the erupting flux rope, but it is quenched inside the flux rope itself because in this region temperatures are low and subject to the broadening technique used to alleviate strong gradients in the transition region (see Section 2).  The ohmic heating dominates the current sheet region at these times (see also Figure\,\ref{heat_along_line.fig}).  At these times, the adiabatic term is positive in the lower part of the current sheet, due to compression of the plasma by the reconnection outflow jet. Conduction is a dominant factor that acts counter to the adiabatic and ohmic heating terms.  It is well-known that thermal conduction will produce a ``thermal halo'' around a current sheet, spreading the temperature away from areas of strong current \citep{YokoyamaShibata1997,Seaton2009,Reeves2010}.  Confirmation of this effect can be seen in the hot temperature structure in Figure \ref{cutplanes_fields.fig}, which is much wider than the area of strong current at the same location.  For the sake of simplicity, we don't show the advection term in Figure \ref{cutplanes_heat.fig}, but it serves mostly as a cooling term in the region of the current sheet, due to the flows and the thermal gradient in this region.

After $t=173\tau_A$, when the dense plasmoid structures begin to be visible in the current sheet region, the adiabatic and thermal conduction terms break up into smaller structures there.  Ohmic heating is still important at $t=173.126\tau_A$, but it is lower in magnitude than the adiabatic heating at the same location.  By $t=175.251\tau_A$ the ohmic heating has faded in the current sheet region, but the adiabatic and conduction terms remain significant. The coronal heating term is strongest in the closed loops that make up the active region, but plays only a minor role in the current sheet region at the times shown.
The area of strong radiation has increased significantly at $t=175.251\tau_A$ in the region where the flare loops should be, outlining the rapid radiative cooling of hot plasma that has been evaporated into the flare loops.

Figure \ref{heat_along_line.fig} plots the heating terms from Equation (\ref{dTdt.eq}) together
with the temperature, density, $v_r$, and $|j|$ along the dashed line shown in Figures \ref{cutplanes_fields.fig} and \ref{cutplanes_heat.fig}.  We also plot the viscous heating due to the numerical dissipation.  Locations where there are stagnation points in the radial velocity are indicated by dashed vertical lines.  Grey vertical lines indicate the ends of the current sheet, defined by a minimum in $|j|$ at either end, as in \citet{Reeves2010}.  

At $t=171.564\tau_A$, Figure \ref{heat_along_line.fig} shows one stagnation point dividing the two outflowing reconnection jets, which is also clear from looking at the radial velocity term plotted in the lower left of Figure \ref{cutplanes_fields.fig}.  The current sheet is short, with the stagnation point near the bottom.  The flux rope that becomes the CME is visible as an enhancement in the density that peaks at about 0.11 R$_s$ along the line.   There are elevated temperatures between 0.02 and 0.10 R$_s$, with the highest temperatures below the current sheet.  

The $dT/dt$ term is positive throughout the current sheet, and highest above and below it, indicating that heating is occurring in those locations.  Below the current sheet, ohmic dissipation and the coronal heating term are contributing in approximately equal quantities to the heating of the plasma in region where flare loops are expected to form.   The adiabatic term also contributes to heating below the current sheet as the radial flows are compressed as they impinge on closed field below.  On the other hand, the adiabatic term cools the plasma at and above the stagnation point as the flows expand there.  Advection removes heat from the region of the stagnation point as hot plasma flows away from it.  Thermal conduction removes heat from the current sheet region and deposits it above the stagnation point, just below the flux rope.  Radiation removes heat from the cool, dense plasma beneath the current sheet, and cools the regions around the dense flux rope.  

\begin{figure*}
\includegraphics[scale=0.87]{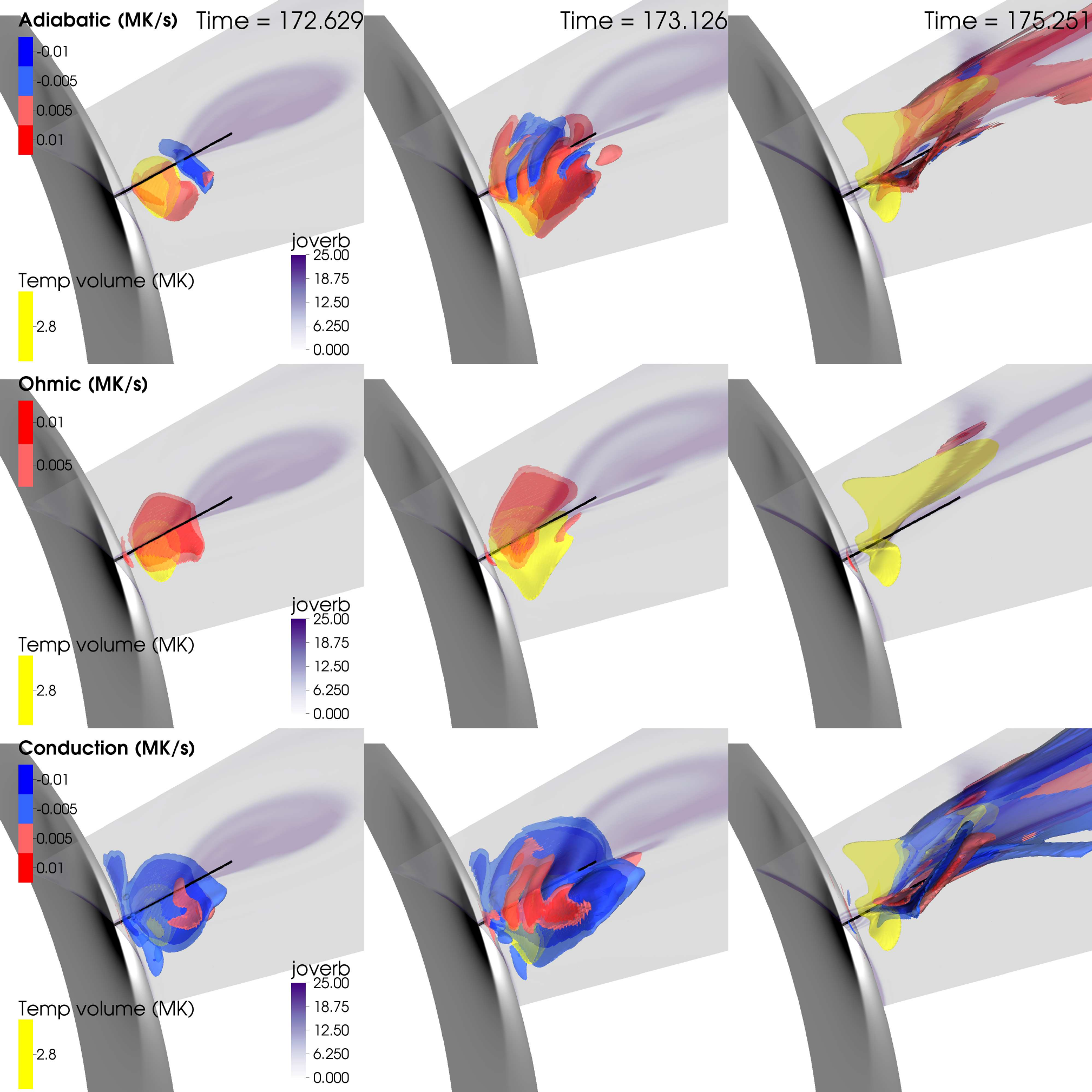}
\caption{\label{volume_side.fig}    Volume contours (red/blue) of the adiabatic term, the ohmic heating term, and the thermal conduction term, in a view with the active region rotated to the west limb, for the last three times shown in Figures \ref{cutplanes_fields.fig}--\ref{heat_along_line.fig}.   The field of view is the same as the box on the images in Figure \ref{cutplanes_big.fig}.  The plane cutting through the active region is the same plane as shown in Figures \ref{cutplanes_fields.fig} and \ref{cutplanes_heat.fig}.  Values of $|J|/|B|$ are shown in this plane.  The black line indicates the line used to plot values in Figure \ref{heat_along_line.fig}.  Also shown is a iso-surface (yellow) of temperature at 2.8 MK. (An animation of this figure is available.)}
\end{figure*}

At $t=172.629\tau_A$, the erupting flux rope has moved higher and the current sheet has become longer.  There is still only one stagnation point in the radial velocity separating the upwards and downwards reconnection jets.  The current has increased, and consequently the heating from the ohmic dissipation has also increased, and it is now significant in the current sheet region.  The adiabatic, advection, conduction, and radiation terms behave similarly as in the previous time shown in Figure \ref{heat_along_line.fig}, but they are more pronounced.  

At $t=173.126\tau_A$ the current sheet has started to break up, possibly due to a plasmoid instability \citep[e.g.][]{Riley2007,Shen2011}. The numerical diffusion starts to be significant at this time due to a combination of a large $\nu_N$ and large stresses from shear flows caused by the instabilities (i.e. the rate of strain tensor $e_{ij}$ in equation \ref{visc_heat.eq} is large), so quantitative values during this phase should be taken with a grain of salt.  However, qualitatively, the evolution of the physical quantities and the heating terms make physical sense. The temperature and density structures in the current sheet region are less smooth than at previous times, and there are now multiple stagnation points in the radial velocity along the current sheet.  The first stagnation point, at 0.06 R$_s$, is caused by a diverging flow.  There is an enhancement in the density structure just below the stagnation point, which is one of the dense plasmoids in the current sheet seen in Figure \ref{cutplanes_fields.fig}.  The temperature is at a local minimum at this point.  The second stagnation point, at 0.08 R$_s$ is due to a converging flow.  There is a local maximum in both the density and the temperature at about the same location, so the converging flow is bringing two plasmoids together.  The final stagnation point is at 0.095 R$_s$, and like the first one, it is a diverging flow.  

The $dT/dt$ term is positive and relatively strong from the flare loops all the way up through the current sheet, indicating strong heating during this phase that causes the plasma to reach a maximum temperature of $\sim$5 MK at t=173.438$\tau_A$. The adiabatic term is the dominant heating term at the converging flow stagnation point and above and below the current sheet, where the reconnection flows impinge on the erupting flux rope and the reconnected flare loops, respectively.   The adiabatic term is diminished at the diverging flows, particularly the lower one at 0.06 R$_s$, where the ohmic heating is the dominant heating term.  From Figure \ref{cutplanes_heat.fig}, we see that there are regions of enhanced adiabatic cooling outside the current sheet at the locations of the diverging flows, but the ohmic heating is confined to the current sheet itself, indicating that the adiabatic term will have effects outside the current sheet whereas the ohmic heating term will not.  We note that even though there are diverging radial flows, the adiabatic term remains mostly positive along a thin region underlying the line in Figure \ref{cutplanes_heat.fig} because of the strong velocity gradient there due to the converging reconnection inflows perpendicular to $v_r$.  The line plots in Figure \ref{heat_along_line.fig} show that thermal conduction is working to smooth out the temperature structure in the current sheet, adding heat at the temperature minima and subtracting it at temperature maxima, and it resembles a mirror image of the adiabatic term because of that term's large heating inputs.  The advection term is removing heat along most of the current sheet, and adding it at the bottom and top of current sheet due to the reconnection flows.  

At  $t=175.251\tau_A$, there are two stagnation points, a lower one with a diverging flow, and an upper one with a converging flow.  The current sheet is quite long at this time, but the lower tip is still relatively low in the corona, consistent with recent theoretical calculations from 2D models \citep{Forbes2018}.  The $dT/dt$ term shows that the upper part of the current sheet is still being heated a small amount, and the adiabatic term is the dominant heating term there.  Below the lower stagnation point there is also some heating from the adiabatic term and a small amount of ohmic heating.  Conduction removes heat from the current sheet.  Below the current sheet, the flare loops are cooling, dominated by conduction and radiation.

\begin{figure*}
\includegraphics[scale=0.87]{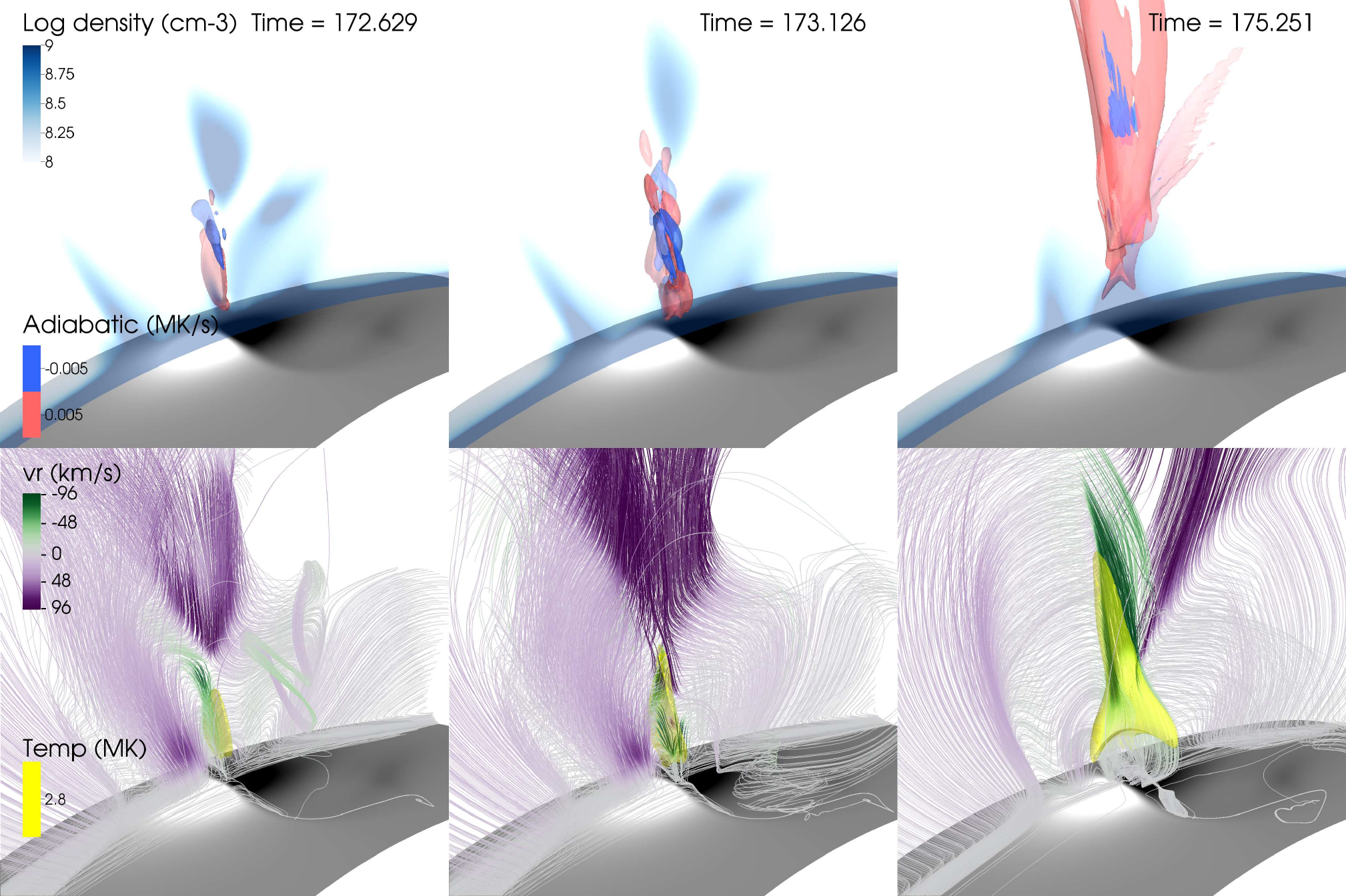}
\caption{\label{vel_stream_lines.fig} Top row:  Volume contours (red/blue) of the adiabatic term in a view looking from the north, parallel to the polarity inversion line, for same three times shown in Figure \ref{volume_side.fig}.  The plane running through the active region is the same as the cut planes used in Figures \ref{cutplanes_fields.fig} and \ref{cutplanes_heat.fig}, and shows values of log density. Bottom row: Selected velocity streamlines for each of the three times, colored by radial velocity, $v_r$. Also shown is a volume contour (yellow) of temperature at 2.8 MK.}
\end{figure*}

\subsection{3D distribution of important energetics terms}

In order to gain insight into the 3D extent and importance of various heating (and cooling) terms, we plot iso-surfaces of these terms at 0.005 and 0.01 MK s$^{-1}$ (as well as at -0.005 and -0.01 MK s$^{-1}$, where appropriate) in Figure \ref{volume_side.fig}.  Also plotted is an iso-surface of temperature at T=2.8 MK, shown in yellow, representing some of the highest temperatures in the simulation.   The black line in Figure \ref{volume_side.fig} is in the same location as the purple dashed line in Figures \ref{cutplanes_fields.fig} and \ref{cutplanes_heat.fig}.  From Figure \ref{volume_side.fig}, it is clear that most of the heat transport occurs near the center of the eruption early in the event, justifying the use of the cut planes in Figure \ref{cutplanes_heat.fig}.  However, there is some heating that occurs outside of the cut plane region, especially at later times.

The first column of Figure \ref{volume_side.fig} shows the iso-surfaces of the selected heating terms at $t=172.629\tau_A$.   At this time, the ohmic dissipation term dominates the heating, and the yellow hot-temperature iso-surface is mostly encompassed by the dark red ohmic heating iso-surface at 0.01 MK s$^{-1}$.  The thermal conduction iso-surfaces at this time show that there is cooling flanked by two areas of heating (light red patches visible on either side of the dark blue patch), indicating that this term is mostly removing heat from the hot current sheet and depositing it nearby, causing the ``thermal halo'' effect.  The iso-surfaces of the adiabatic term at this time show a region of heating (mostly light red, 0.005 MK s$^{-1}$) low in the eruption, coincident with the 2.8 MK temperature iso-surface, as well as a slightly smaller region of cooling (light and dark blue) directly above the main heating region.  There also is a small separated iso-surface of adiabatic heating at 0.005 MK s$^{-1}$ at about the same height as the blue iso-surfaces, and to the south of the cut plane. 

The top row of Figure \ref{vel_stream_lines.fig} shows the adiabatic heating iso-surfaces from a different viewpoint that is more parallel with the polarity inversion line, along with an image of density in the cut plane.   To illustrate the flows in the reconnection region, we have plotted velocity streamlines in the bottom row of Figure \ref{vel_stream_lines.fig}, colored by their value of $v_r$. At $t=172.629\tau_A$, the flow pattern is relatively simple, with upflows (purple) and downflows (green) emanating from a stagnation point above the polarity inversion line. The iso-surface of adiabatic heating at this time is due to the compression from reconnection outflows (green vertical lines) impinging upon closed loops below as well as compression from the reconnection inflows (grey horizontal lines).  The separated light red iso-surface of adiabatic heating higher up is co-located with some of the strongest radial outflows, and the heating there is due to these large reconnection outflows impinging upon the bottom of the flux rope and compressing plasma there.  The blue iso-surfaces representing adiabatic cooling are located at the stagnation point above the polarity inversion line, where the reconnection outflows diverge.  

\begin{figure*}
\includegraphics[scale=0.6]{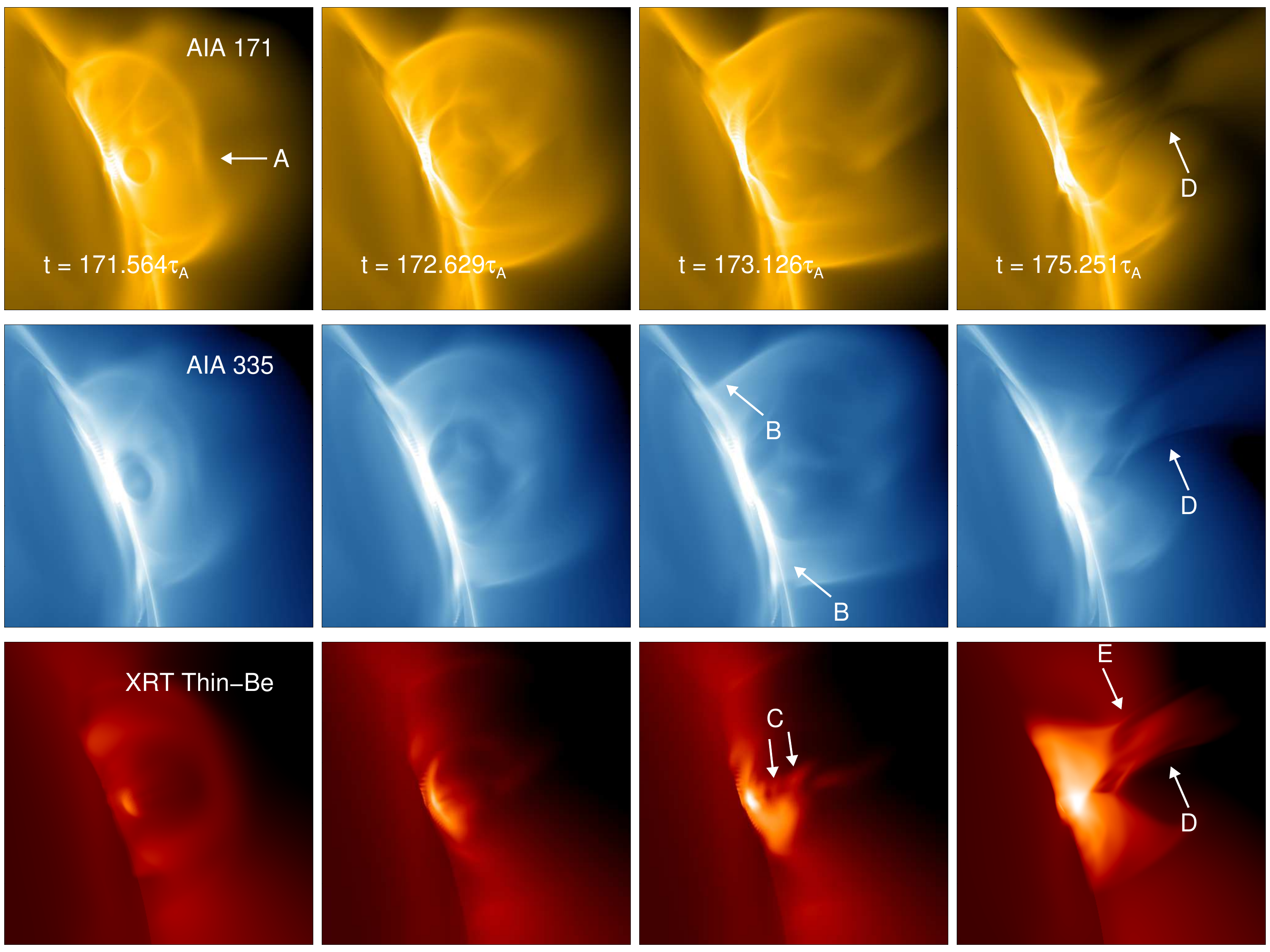}
\caption{\label{simulated_emission.fig} Simulated emission for the AIA 171 \AA\ (top row), AIA 335 \AA\ (middle row) and XRT Thin-Be (bottom row) channels for 
the same times as shown in Figures \ref{cutplanes_fields.fig}--\ref{heat_along_line.fig}. Location `A' marks the top of the
flux rope, location `B' marks the legs of the erupting flux rope, location `C' marks the blobs that form during the eruption, location `D' marks a bright, high supra-arcade structure between the flare loops in the XRT image, and location `E' marks a narrow, thin structure extending from the flare loops.  The active region has been rotated to the limb, and the field of view is the same as that shown in Figure \ref{volume_side.fig}. (An animation of this figure is available.)}
\end{figure*}

The second column in Figure \ref{volume_side.fig} shows the selected heating terms at $t=173.126\tau_A$, when the plasmoids start to form in the current sheet region.  At this time, the ohmic heating iso-surfaces overlap with the 2.8 MK temperature iso-surface, but they are not completely co-spatial with the high temperatures, with an offset slightly to the north.  Alternating red and blue contours are seen in the adiabatic term around where the plasmoids form along with the multiple stagnation points.  The thermal conduction term also shows alternating red and blue iso-surfaces in roughly opposite locations to the ones in the adiabatic term, indicating that the conduction term is removing heat where the adiabatic term is adding it.   The top middle panel of Figure \ref{vel_stream_lines.fig} shows the alternate view of the adiabatic iso-surfaces at this time, indicating that the strong heating remains located in the center of the active region, above the polarity inversion line.  The bottom middle panel of Figure \ref{vel_stream_lines.fig} shows that the flow field at this location has become more complex, with upflows and downflows intermingled.  

At the final time shown in Figure \ref{volume_side.fig}, $t=175.251\tau_A$, the ohmic heating iso-surface and the hot temperature iso-surface hardly overlap at all, indicating that the ohmic dissipation in the current sheet is not the primary contributor to the hot plasma in the eruption at this time.  The ohmic term is also smaller in both magnitude (light red, 0.005 MK s$^{-1}$) and spatial extent than the other terms shown.  The adiabatic heating is now strongest outside of the cut plane, and there is also an extension of the adiabatic heating iso-surface pointing out of the page to the south of the cut plane in the view in Figure \ref{volume_side.fig}.  The alternate view of the adiabatic iso-surfaces in the top right panel of Figure \ref{vel_stream_lines.fig} shows this geometry more clearly, with the northern part of the active region in the foreground and the southern part in the background.   Similar to the other times shown, the thermal conduction term shown in the lower right panel of Figure \ref{volume_side.fig} is cooling in regions where the adiabatic term is heating, and it is heating the plasma surrounding the cooling regions.  This effect is most easily seen in the iso-surfaces extending out of the page south of the cut plane, where a blue iso-surface can be seen sandwiched between two red ones.

The temperature iso-surface at T=2.8 MK at $t=175.251\tau_A$ shows that high temperatures reach high altitudes to the north of the active region center.  At this location, the iso-surfaces of the adiabatic term at 0.005 and 0.01 MK s$^{-1}$ surround the yellow high temperature iso-surface.   The formation of this region of high temperatures corresponds in time to the rotation of the current sheet structure shown in Figure \ref{flux_rope.fig} and the accompanying movie.  As this rotation progresses, the inverse S-shaped ends of the current sheet structure become more tightly curled.   The strong temperature increase on the north side of the active region is due to the reconnection inflows and downflows compressing the plasma as the magnetic structures rotate and curl inward.  This effect can be seen clearly in the plot of the velocity streamlines at this time, shown in the bottom right panel of Figure \ref{vel_stream_lines.fig}.  Grey-colored horizontal stream lines bend into green-colored (i.e. downflowing) streamlines at the location of the red adiabatic iso-surface.  A similar effect exists in the southern part of the active region, but because of the asymmetries in the active region field, the current sheet bends over and the adiabatic compression stays lower in height than in the north.  It is also notable that the stagnation point is much lower in the southern part of the active region, so that down flows are confined to lower heights and only upflowing (purple) velocities are visible from the viewpoint shown in Figure  \ref{vel_stream_lines.fig}.

\begin{figure*}
\includegraphics[scale=0.50]{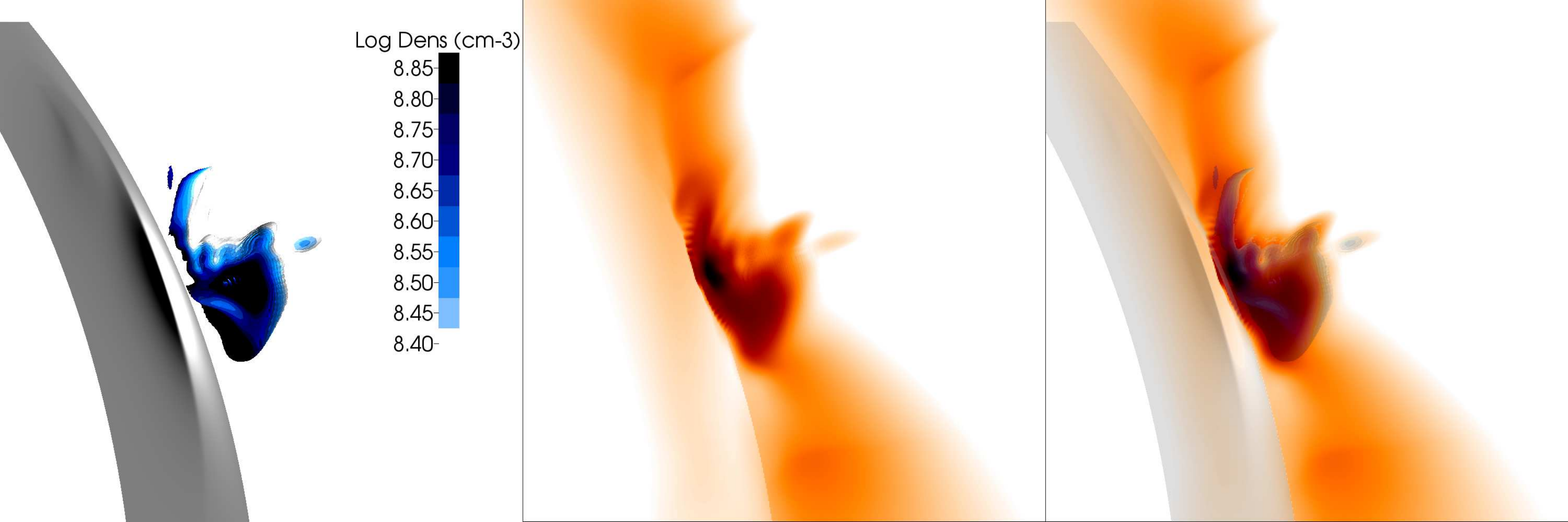}
\caption{\label{blob_emission.fig}  Left panel: Density contours for plasma with temperatures of 2 - 3.5 MK for time $t=173.126\tau_A$.  Middle panel: Simulated XRT Be thin image with an inverted color scale for the same time.  Right panel: Overlay of the two panels.}
\end{figure*}

\section{EUV and X-ray emission}

We simulate the emission from this model in bandpasses from the X-Ray Telescope \citep[XRT;][]{Golub2007} on {\it Hinode} and the Atmospheric Imaging Assembly \citep[AIA;][]{Lemen2011} on the {\it Solar Dynamics Observatory}.  The emission can be calculated using the equation
\begin{equation}
I = \int n^2_e(l)f_i(T(l),n_e(l))dl 
\end{equation}
where $n_e$ is the electron density, $T$ is the temperature, $f_i$ is the instrument response function, and $l$ is the line of sight.  The instrument response functions for XRT and AIA are provided in the SolarSoftware \citep[SSW][]{Freeland1998} IDL distribution.  The results are shown in Figure \ref{simulated_emission.fig} and the accompanying movie.  We show the simulated AIA 335 \AA\ emission because its temperature response (which peaks around 2.5 MK) is the best suited to show the hottest features ($\sim$ 3 MK) in the eruption.  The XRT Thin-Be filter also shows the hot features nicely because of its broad temperature coverage.  We also show the AIA 171 \AA\ emission for comparison with a cool channel.

In the initial panels in Figure \ref{simulated_emission.fig}, the most prominent feature is a sheath of material at coronal temperatures that surrounds the cooler flux rope (marked as `A' in Figure \ref{simulated_emission.fig}).  As the eruption progresses, the legs of the CME (marked as `B') are visible in the AIA channels, though this emission dissipates by the time shown in the last column of Figure \ref{simulated_emission.fig}.  The simulated XRT images clearly show the hot structures associated with the flare below the erupting flux rope.  The XRT image at $t=173.126\tau_A$ also shows some bright blobs in intensity, marked as location `C'.  The movie accompanying Figure \ref{simulated_emission.fig} shows these blobs traveling both up and down in the XRT and AIA 335 \AA\ images between $t=173.126\tau_A$ and $t=175.251\tau_A$.  Additionally, there are some bright blobs that are seen traveling upward in both of the AIA filters late in the movie. In the final column, the XRT image shows a structure in between the flare loops that reaches higher in the corona than the other flare structures (location `D').  This structure is faintly visible in the AIA 335 \AA\ image, and the AIA 171 \AA\ image shows a decrease in intensity at this location (see arrows in the last column of Figure \ref{simulated_emission.fig}).  There is also a long and narrow bright structure emanating from the northernmost set of flare loops seen in the XRT image at this time (location `E').  Similar structures have often been identified as the flare current sheet by observers, and have been used to estimate the properties of the current sheet such as its width \citep[e.g.][]{Warren2018,Yan2018,Savage2010}.

The middle panel in Figure \ref{blob_emission.fig} shows the XRT image at $t=173.126\tau_A$ with an inverted color scale to highlight the location of the blobs. At this time current sheet starts to break up, and the intensity blobs seen in the simulated XRT image are due to dense plasmoid-like structures forming in the current sheet. The left panel of Figure \ref{blob_emission.fig} shows volume contours of density for plasma with temperatures restricted to the range of 2-3.5 MK.  The right panel is an overlay of these two images, showing that the intensity enhancements in the XRT image correspond to the areas of enhanced density. The blobs are not visible in the AIA 335 \AA\ because they are obscured by cooler structures (visible in the 171 \AA\ image as well) along the line of sight.  These structures are too cool to be seen in the XRT passband, so the blobs show most clearly in the XRT intensity image. 

\begin{figure*}
\includegraphics[scale=0.5]{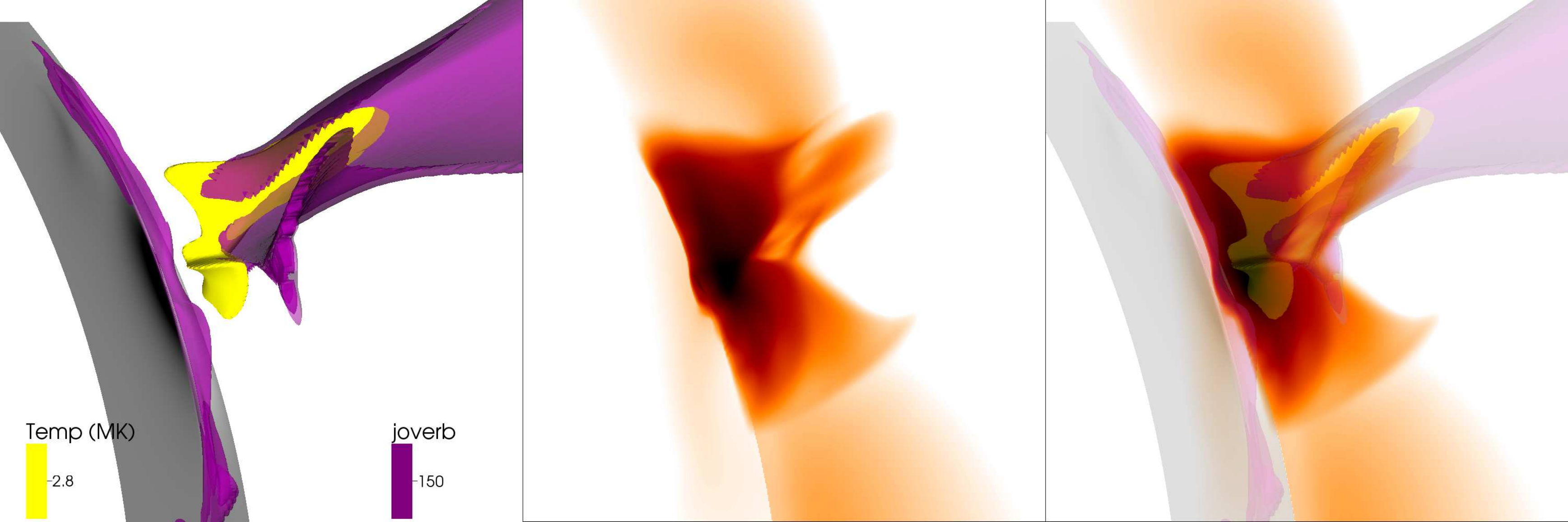}
\caption{\label{joverb_emission.fig}  Left panel: Isovolume contours $|J|/|B|$ (purple) and the temperature (yellow)  for time $t=175.251\tau_A$.  Middle panel: Simulated XRT Be thin image with an inverted color scale for the same time.  Right panel: Overlay of the two panels. }
\end{figure*}

The middle panel in Figure \ref{joverb_emission.fig} shows an inverted color scale XRT image late in the eruption, at $t=175.251\tau_A$.  In this image, the long, thin structure at the top of the northern set of flare loops (labeled `E' in Figure \ref{simulated_emission.fig}) is quite noticeable. The left panel of the figure shows a temperature iso-surface at T=2.8 MK (yellow) along with an iso-surface of $|J|/|B|$ (purple), indicating the location of the current sheet.  Note that the $|J|/|B|$ iso-surface has a similar structure to that of the iso-surface of the adiabatic compression (red) in the upper right panel of Figure \ref{volume_side.fig}, complete with a structure that extends out of the page in the south.  The right panel shows an overlay of these two images.  From this image, it can be seen that the tall structure labeled `D' in Figure \ref{simulated_emission.fig} overlaps with the volume contour of temperature at 2.8 MK.   The long, thin spire of XRT emission at the top of the northern loops (labeled `E' in Figure \ref{simulated_emission.fig}) is aligned with one edge of the current sheet, where it folds over on itself.  However, the extent of the current sheet, which is a 3D structure that curves around the yellow high temperature iso-surface, is not evident in the XRT emission.  Therefore long, thin emission features do not necessarily outline the entirety of the current sheet, and should be treated with caution in observations.

\section{Discussion and Conclusions \label{discussion.sec}}

In this paper, we analyze plasma heating in the flare current sheet of a simulated solar eruption and 
we produce synthetic XRT and AIA images to understand how the structures around the current sheet manifest in satellite observations.  
Our main findings are:

\begin{enumerate}

\item Ohmic heating is an important contributor to plasma heating in the current-sheet region early in the eruption. After the onset of the tearing instability and plasmoid formation, adiabatic compression becomes the dominant heating mechanism;

\item Thermal conduction transports thermal energy away from the current-sheet region throughout the reconnection process, widening the region of high temperatures;

\item Simulated XRT emission shows a faint, high-altitude structure above the flare loops (location `D' in Figure \ref{simulated_emission.fig}) in the late phase of the eruption that corresponds to plasma heated through adiabatic compression.  This structure is equivalent to the supra-arcade plasma sheet sometimes seen in observations of long duration events;

\item The thin feature seen in the simulated XRT emission at the top of cusp-shaped loops (location `E' in Figure \ref{simulated_emission.fig}) may be interpreted to outline the location and orientation of the flare current sheet. However, while this feature is indeed co-spatial with one edge of the highly curved current sheet, it outlines only a very small fraction of it.

\end{enumerate}
 
Early in the simulated eruption, the ohmic dissipation is the dominant heating mechanism in the current sheet region.  The strongest heating due to the ohmic dissipation occurs in the center of the active region, along the spine of the inverse S-shaped region of the current sheet.  In the middle stages of the eruption, the current sheet decreases in strength in the center of the active region and increases in the northern and southern 
regions, where it wraps around the legs of the CME flux rope.  At this time, the location of the strongest ohmic dissipation has moved northward, indicating that the currents are somewhat stronger around the northern leg, which is likely a consequence of the asymmetric eruption. In the later phase of the eruption, the ohmic dissipation 
decreases and becomes insignificant compared to the heating caused by the adiabatic compression.

The current sheet remains intact in the early phase of the simulated eruption, but later on it breaks up, presumably due to the plasmoid instability, and a few plasmoid-like features are seen. Once the plasmoids form, heating from adiabatic compression becomes important, as shown in Figure \ref{volume_side.fig}.  Simulations of reconnection \citep[e.g.][]{Bhattacharjee2009,Nishida2009} and laboratory reconnection experiments \citep[e.g.][]{JaraAlmonte2016} have shown that the formation of plasmoids increases the reconnection rate, and hence the inflow velocity to the current sheet. Larger inflows lead to larger velocity gradients, since the inflow velocity goes to zero at the center of the current sheet. The onset of the plasmoid instability therefore increases the heating of the surrounding plasma by adiabatic compression as the inflow velocity increases. We find observational signatures for the plasmoids in the form of bright blobs that are visible in the simulated XRT emission (see Figure \ref{blob_emission.fig}).  

The strong adiabatic heating in the late phase of the eruption is similar to the effect described in \cite{Birn2009}, where adiabatic compression causes heating in layers along the current sheet. In our simulation, heat is transported away from the current sheet also by thermal conduction, which was not included in the simulations by \cite{Birn2009}. We find that elevated temperatures above the flare loops are only seen in the northern part of the active region, however, because strong reconnection outflows in the current sheet in the southern part of the active region allow the heat to be advected upward, away from the loops, similar to the effect seen in the advection term in Figure \ref{heat_along_line.fig} at $t=173.136\tau_A$.  

The sustained heating due to adiabatic compression late in the event at $t=175.251\tau_A$ leads to hot ($\sim$3 MK) plasma that manifests as supra-arcade emission in the XRT Be-Thin filter (see the feature marked `D' in Figure \ref{simulated_emission.fig}). This emission feature is less pronounced in the AIA 335 \AA\ and 171 \AA\ channels, and completely absent in the AIA 131 \AA\ and 94 \AA\ channels (not shown), which are sensitive to hotter temperatures.  Another observational manifestation of the adiabatic compression near the current sheet is a bright, linear feature in the simulated XRT emission (see the feature marked `E' in Figure \ref{simulated_emission.fig}).   This emission is not from particularly hot plasma, but rather from plasma that has been compressed by reconnection inflows up against the current sheet. Because the plasma is optically thin, the emission is enhanced in places where the current sheet bends around itself.  The bright, linear emission feature formed by this overlap is co-spatial with the current sheet (Figure \ref{joverb_emission.fig}), but it is not indicative of the entire extent of the current sheet.  Rather, this feature is formed due to a combination of adiabatic compression around the current sheet and line-of-sight effects conspiring to form a thin feature of emission.  This result indicates that real 3D flare current sheets can be complex, and that emission images may outline only a small fraction of their total extent.
This finding shows that linear features observed in observations should not necessarily be assumed to encompass the entire current sheet, and the 3D geometry of the erupting region should be considered.

Hot supra-arcade plasma sheets are often observed in the late phase of long-duration solar flares \citep[e.g.][]{McKenzie1999,Innes_b2003,Reeves2011,Savage2012,Hanneman2014,Reeves2017}, and these structures persist for much longer than the conductive cooling time \citep{Reeves2017}. This observation indicates that either some continual heating is occurring, or that thermal conduction is somehow suppressed. Based on the simulation results described here, we suggest that there may be additional heating in regions where supra-arcade plasma sheets are observed, due to adiabatic compression as fast reconnection continues in the current sheet late in the eruption. This mechanism would be consistent with spectroscopic observations that show that post-eruption current sheet regions have coronal abundances \citep{Landi2012,Warren2018} rather than photospheric abundances, which is what would be expected in a chromospheric evaporation type mechanism. 

It is worth noting that we have assumed ionization equilibrium for the calculation of our synthetic emission images. Non-equilibrium effects could change the intensities to some degree, especially in the supra-arcade region where the plasma is less dense than in the flare loops. Results obtained from a 2D thermodynamic MHD CME simulation have shown that non-equilibrium ionization effects can cause emission to be underestimated low down in the current sheet region, and to be overestimated higher up \citep{Shen2013ApJ}.  We plan to examine the effects of non-equilibrium ionization on the emission properties in CME simulations in the future by using an in-line calculation of ionization states during the solution of the MHD equations, a feature that has been implemented in the MAS model \citep{Lionello2019}.  This exercise will also be useful for modeling {\it in situ} elemental and composition measurements of flux rope ICMEs \citep[e.g.][]{Lepri2010,Reinard2012,Rivera2019}, an endeavor previously only attempted with {\it post facto} ionization calculations in 2D simulations \citep{Lynch2011}, or highly idealized 3D configurations \citep{Rakowski2011}.

The eruption studied here is not very energetic, so the maximum plasma temperatures ($\sim$3-5 MK) are not very high with respect to typical flare temperatures, which have been observed to reach tens of MK. We plan to apply a similar analysis to more energetic simulated eruptions in the future, {\citep[such as the one presented in][]{Torok2018}.  We note that MHD simulations cannot capture the contributions of accelerated particles to the energy dissipation and transport, which are important during the impulsive phase of large flares \citep{Emslie2005,Aschwanden2017}.  Despite this restriction, simulations like the one presented here can provide very useful information on the mechanisms and locations of thermal energy release and transport in solar eruptions.

\section*{Acknowledgements} 
The authors would like to thank the anonymous referee for comments that improved this paper.  KKR would like to thank Cooper Downs for his help with some of the MAS diagnostic tools.  The work of KKR for this project is supported by the NSF SHINE program, grants AGS-1156076, AGS-1460165, and AGS-1723425. NAM acknowledges support from the NSF SHINE program, grant AGS-1156076 and the U.S. Department of Energy, Office of Science, Fusion Energy Sciences program under Award Number DE-SC0016363 awarded through the NSF-DOE Partnership in Basic Plasma Science and Engineering.    
TT was supported by NASA's LWS program (Award Number NNH13ZDA001N), and JL and ZM acknowledge support from AFOSR (contract FA9550-15-C-0001) and NASA LWS (Award Number NNH14CK98C).  Computational resources were provided by the NSF XSEDE program (at the Texas Advanced Computing Center and the San Diego Supercomputer Center) and the NASA Advanced Supercomputing Division at Ames Research Center. {\it Hinode} is a Japanese mission developed and launched by ISAS/JAXA, with NAOJ as domestic partner and NASA and STFC (UK) as international partners. It is operated by these agencies in co-operation with ESA and NSC (Norway). This work has benefited from the use of NASA's Astrophysics Data System.

\listofchanges
\end{document}